\documentclass[]{pasj02} 
\usepackage{anyfontsize}
\usepackage{mathtools}
\usepackage[switch,mathlines]{lineno} 
\usepackage{bm}
\usepackage{diagbox}
\usepackage{pdflscape}

\jyear{2024}
\Received{2023/11/03}
\Accepted{2024/12/03}

\begin{document} 

\title{Smart Kanata: A Framework for Autonomous Decision Making in Rapid Follow-up Observations of Cataclysmic Variables}

\author{
Makoto \textsc{Uemura},\altaffilmark{1}\altemailmark\orcid{0000-0002-7375-7405} \email{uemuram@hiroshima-u.ac.jp}
Yuzuki \textsc{Koga},\altaffilmark{2}
Ryosuke \textsc{Sazaki},\altaffilmark{2}
Tomoya \textsc{Yukino},\altaffilmark{3,4}
Tatsuya \textsc{Nakaoka},\altaffilmark{1}\\
Ryo \textsc{Imazawa},\altaffilmark{2}
Taichi \textsc{Kato},\altaffilmark{5}
Daisaku \textsc{Nogami},\altaffilmark{5}
Keisuke \textsc{Isogai},\altaffilmark{5,6}
Naoto \textsc{Kojiguchi},\altaffilmark{5}\\
Kenta \textsc{Taguchi},\altaffilmark{5}
Yusuke \textsc{Tampo},\altaffilmark{5,7,8}
Hiroyuki \textsc{Maehara},\altaffilmark{9}
and
Shiro \textsc{Ikeda}\altaffilmark{10}
}
\altaffiltext{1}{Hiroshima Astrophysical Science Center, Hiroshima University, Kagamiyama 1-3-1, Higashi-Hiroshima, Hiroshima 739-8526, Japan}
\altaffiltext{2}{Department of Physics, Graduate School of Advanced Science and Engineering, Hiroshima University, 1-3-1 Kagamiyama, Higashi-Hiroshima, Hiroshima 739-8526, Japan}
\altaffiltext{3}{Institute of Astronomy, Graduate School of Science, The University of Tokyo, 2-21-1 Osawa, Mitaka, Tokyo 181-0015, Japan}
\altaffiltext{4}{Department of Astronomy, Graduate School of Science, The University of Tokyo, Hongo 7-3-1, Bunkyo-ku, Tokyo 113-0033, Japan}
\altaffiltext{5}{Department of Astronomy, Kyoto University, Kitashirakawa-Oiwake-cho, Sakyo-ku, Kyoto, Kyoto 606-8502, Japan}
\altaffiltext{6}{Department of Multi-Disciplinary Sciences, Graduate School of Arts and Sciences, The University of Tokyo, 3-8-1 Komaba, Meguro, Tokyo 153-8902, Japan}
\altaffiltext{7}{South African Astronomical Observatory, PO Box 9, Observatory, 7935, Cape Town, South Africa}
\altaffiltext{8}{Department of Astronomy, University of Cape Town, Private Bag X3, Rondebosch 7701, South Africa}
\altaffiltext{9}{Subaru Telescope Okayama Branch Office, National Astronomical Observatory of Japan, National Institutes of Natural Sciences, 3037-5 Honjo, Kamogata-cho, Asakuchi, Okayama 719-0232, Japan}
\altaffiltext{10}{The Institute of Statistical Mathematics, 10-3 Midori-cho, Tachikawa, Tokyo 190-8562, Japan}



\maketitle

\begin{abstract}
Studying the early stages of transient events provides crucial information 
about the fundamental physical processes in cataclysmic variables (CVs). 
However, determining an appropriate observation mode immediately after the 
discovery of a new transient presents challenges due to significant 
uncertainties regarding its nature. We developed a framework designed for autonomous decision making in prompt follow-up observations of CVs using the 
Kanata 1.5-m telescope. The system, named Smart Kanata, first estimates the 
class probabilities of variable star types using a generative model. It then 
selects the optimal observation mode from three possible options based on the 
mutual information calculated from the class probabilities. We have operated 
the system for $\sim 300$ days and obtained 21 samples, among which automated 
observations were successfully performed for a nova and a microlensing event. 
In the time-series spectra of the nova V4370 Oph, we detected a rapid 
deepening of the absorption component of the H$\alpha$ line. These initial 
results demonstrate the capability of Smart Kanata in facilitating rapid 
observations and improving our understanding of outbursts and eruptions of 
CVs and other galactic transients.
\end{abstract}


\section{Introduction}

Cataclysmic variables (CVs) are close binary systems in which a white dwarf
experiences accretion from a companion star that fills its Roche lobe
(for a review, see \cite{war95book}). This accretion process results in 
transient phenomena, including nova eruptions and dwarf nova outbursts. 
Uninterrupted observations spanning the entire duration of these phenomena 
are essential to understand them. However, capturing the initial phase, 
particularly the rapidly rising phase, presents challenges due to its short 
time-scale of less than 1 day. Exploring the initial stage of transients, 
typically characterized by an unsteady transitional state, provides valuable 
insights into their underlying physical models.

In novae, the accretion triggers a runaway thermonuclear burst on the surface 
of the white dwarf. This burst leads to the ejection of accumulated matter, 
making a photosphere that envelops the entire binary star system (for a 
recent review, see \cite{del20nova}). The dynamic formation and subsequent 
evolution of the ejecta have recently been investigated through 
spectroscopic observations conducted during the initial stages of nova 
eruptions. In the pioneering work of \citet{ara15tpyx}, an optical spectrum 
was acquired within an extremely early temporal window, specifically at 0.19 
days following the initial detection of the nova eruption in T Pyx. This 
spectrum exhibited distinct features reminiscent of Wolf-Rayet stars, 
including N\,\emissiontype{V}, \emissiontype{IV}, C\,\emissiontype{IV} and 
He\;\emissiontype{II} emission lines, which were unprecedented for novae. In 
particular, these highly excited lines disappeared within 2 days. A similar 
spectrum was identified during the rising phase of Gaia22alz 
\citep{ayd23g22alz} and V1405 Cas \citep{tag23v1405cas}. Even in ordinary 
emission lines, \citet{yam10usco} detected the rapid evolution of 
high-velocity components within the first day of the nova eruption in U Sco. 
These observations indicate the presence of previously uncharacterized 
conditions in the early phases of nova eruptions, suggesting that further 
investigation is needed to understand the underlying mechanisms.

Dwarf novae exhibit a repetitive pattern of outbursts that are triggered by
the thermal instability of the accretion disk \citep{osa96dn}. Within the 
category of dwarf novae, WZ Sge-type stars exhibit distinctive features: 
large outburst amplitudes ($\sim 8$\,mag), long recurrence time 
($\sim 10$\, yr), and the emergence of early superhumps during the initial 
week of their outburst \citep{how95toad, kat15wz}.
The early superhumps are short-term periodic 
fluctuations characterized by a doubly-peaked profile of the orbital period. 
They arise from geometric irregularities within the disk structure. However, 
the underlying mechanism responsible for the disk structure remains 
poorly understood. WZ Sge stars have the most extreme binary mass ratio 
($M_2/M_1\lesssim 0.1$) among CVs. Consequently, the accretion disk in 
these systems is expected to approach the 2:1 resonance radius. At 
this radius, tidal dissipation would influence the disk structure 
\citep{osa02esh}. This scenario emphasizes the importance of early superhumps 
for studying the tidal effect in accretion disks in binary systems with 
extreme mass ratios.

In the case of both novae and dwarf novae, it is necessary to observe the
very early stages of the transient event to understand the 
aforementioned phenomena. The appropriate mode of the follow-up observation 
depends on the nature of the transient. For instance, photometric 
observations are employed to identify early superhumps, while spectroscopic 
observations are used to investigate the formation process of the nova 
ejecta. Unlike objects such as supernovae, where spectroscopic observations are consistently prioritized for follow-up, the availability of this diverse selection complicates immediate follow-up observations of CVs.
Determining the suitable observation mode for a new transient is a 
challenging task because of the limited information available, leading to
substantial uncertainties in its nature immediately after its discovery. This 
decision-making process has traditionally been the responsibility of experts 
in CVs; however, there exists a risk of overlooking critical phases of the 
event without the guidance of these experts during nighttime observations.

Another problem arises from the substantial increase in the count of transient
events suitable for potential follow-up observation due to large survey
projects in recent years (e.g. ASAS-SN in \cite{asassn}; Zwicky Transient
Facility in \cite{ztf}). The use of manual decision making by experts
has become constrained as a result of this increase.

In this paper, we present a framework designed for the autonomous 
decision-making process in rapid follow-up observations of CVs using the 
Kanata 1.5-m telescope at the Higashi-Hiroshima Observatory. We call this 
system Smart Kanata. The novelty of our Smart Kanata as an observation system 
is that it automatically determines and executes the observation expected to 
have the largest information gain using the framework of information theory, 
rather than simply automatically executing a predetermined observation.

Previous studies have explored machine learning techniques for classifying 
transient events and variable stars (for a review, see \cite{for21alerce}). 
However, most existing models are designed for broad classifications of 
variable star types or focus exclusively on particular types, such as 
supernovae (e.g., \cite{mut19rapid}) and pulsating variables (e.g., 
\cite{jay19asassn}). In contrast, our study targets finer classifications of 
CVs, including subtypes of dwarf novae, necessitating the development of a 
new and specialized classification model. 

The automation of observations, from the detection of transient events to 
their follow-up, has advanced significantly, particularly in studies of gamma-
ray bursts and supernovae (e.g., \cite{gal11sn}). These automated systems 
have made substantial contributions to our understanding of such phenomena; 
however, they often rely on static decision trees or predefined observation 
priorities. \citet{mah08bayes} and \citet{djo11sss} introduced the concept of 
using probabilistic models for dynamic, autonomous decision-making processes 
to determine the optimal observation mode. Building on this approach, our 
study implements it within an observational system and demonstrates its 
effectiveness in real-time follow-up of CVs.

The structure of this paper is as follows: Section 2 provides a comprehensive 
description of the system. Section 3 presents the initial results observed 
during the 11-month period from December 2023 to October 2024. In Section 
4, we discuss the performance of the system, particularly the classifier, and 
the implications of the acquired data. Finally, Section 5 offers a summary of 
our findings.

\section{The Smart Kanata system}

In Smart Kanata, we classify objects into five categories: nova (class N), 
dwarf nova (class DN), WZ Sge-type dwarf nova (class WZ), mira (class M), and 
flare star (class F). The latter two classes are included because they are 
sometimes mistaken for transient phenomena in CVs. We use eight features for 
each sample: the galactic coordinates ($\ell$ and $b$), the absolute 
magnitudes at maximum brightness ($M_{\rm max}$) and in quiescence 
($M_{\rm qui}$), the transient amplitude ($a$), and the quiescent colors ($g-r$, $r-i$, 
and $i-z$). The goal of Smart Kanata is to select the optimal follow-up 
observation mode for use with the Kanata telescope. We consider three 
available observation modes: multiband imaging to obtain the $B-V$ color 
index (mode $B-V$), spectroscopy to measure the equivalent width of 
${\rm H}\alpha$ (mode EW), and time-series imaging to investigate short-term 
periodic variations (mode Var).

The Smart Kanata system monitors new transients from online services to find  
potential targets suitable for observation using the Kanata telescope 
(\S2.1). Upon detecting a candidate target, Smart Kanata classifies the 
object and selects the optimal observation mode. We introduce the 
framework of these processes in \S2.2. Details about the classifier and the 
decision-making process for observations are described in \S2.3 and \S2.4, 
respectively. Once the decision is made, the system initiates automated 
observations, as explained in \S2.5.

\subsection{Target selection}

The Smart Kanata system monitors online platforms for new transients. These 
platforms include the Transient Name Server 
(TNS)\footnote{https://www.wis-tns.org}, CBAT's "Transient Objects 
Confirmation Page" 
(TOCP)\footnote{http://www.cbat.eps.harvard.edu/unconf/tocp.html}, 
ASAS-SN Transients 
\citep{asassn}, and ANTARES \citep{antares}. Our system identifies transients 
that are brighter than a certain limit ($m_{\rm lim}$) and are detected 
within a defined temporal window ($\Delta t_{\rm dis}$). Detailed parameter
values are provided in Table~\ref{tab:smartk_params}.

After the system identifies transients, it verifies if they correspond to 
either recognized variable stars or active galactic nuclei (AGN) listed in 
the available database. Our search for recognized variable stars involves
referencing the AAVSO VSX catalog \citep{aavso_vsx}, the ATLAS catalog 
\citep{atlas_cat}, and the ASAS-SN catalog \citep{asassn}. For identifying known 
AGN, we consult the WISE AGN candidates \citep{wise_agn}, AGNs detected in 
the mid-infrared (MIR) through AllWISE data \citep{wise_mir_agn}, Blazar 
Radio and Optical Survey \citep{bros}, and the FIRST-NVSS-SDSS AGN sample 
catalog \citep{first}. The system identifies new transients as known sources 
if they fall within a distance of $r_{\rm match}$ from their position. The 
default value for $r_{\rm match}$ is $2\arcsec.0$, as shown in 
Table~\ref{tab:smartk_params}.

We excluded clear supernovae (SNe) from our target candidates. We identify 
such SNe as events whose host galaxy is listed in the online services. In 
addition, we consider an event with a neighboring galaxy to be a potential 
supernova, and exclude it. The neighboring galaxy is identified as a galaxy 
within a distance of $r_{\rm gal}$ from the event's position in the NGC 
2000.0 Catalogue \citep{ngc2000}, SDSS-DR8 galaxies categorized by WND-CHARM 
\citep{sdss_gal}, or the GLADE v2.3 catalog \citep{glade}. The default value 
for $r_{\rm gal}$ is set to $10\arcsec.0$, as shown in 
Table~\ref{tab:smartk_params}. 

The observation sequence is initiated if a new transient event is identified
with a known nova or WZ Sge-type dwarf nova in the AAVSO VSX catalog, as 
detailed in \S~2.4. If the event is not found in the aforementioned catalogs 
and is not identified as a SN, we then initiate the classification and 
subsequent decision-making sequences described in \S~2.2 and 2.3, 
respectively.

\begin{table*}
  \tbl{System parameters of Smart Kanata.}{%
  \begin{tabular}{ccc}
      \hline
      Parameter & Meaning & Value\\ 
      \hline
      $m_{\rm lim}$ & Magnitude limit for faint objects to be extracted & 16.0\\
      $\Delta t_{\rm dis}$ & Elapsed time since discovery to be extracted & $2.0\;{\rm d}$\\
      $r_{\rm match}$ & Radius to search for the counterpart of the object. & $2\arcsec.0$ \\
      $r_{\rm gal}$ & Radius to search for the galaxy near the object & $10\arcsec.0$ \\
      $p(k)$ & Prior probability for the classifier & $(0.2, 0.2, 0.2, 0.2, 0.2)$\\
      $p_{\rm obs}$ & Threshold of class probability to initiate observations & 0.95 \\
      $p_{\rm target} $ & Threshold of class N or WZ probability to initiate observations & 0.10 \\
      \hline
    \end{tabular}}\label{tab:smartk_params}
\end{table*}

\subsection{Uncertainty of the class and choosing the follow-up observation mode}

Smart Kanata uses probabilistic models for the classification of transients 
and the selection of follow-up observation modes. Here, we describe the 
mathematical framework.

Let $\pi(k)$ be the initial guess, or in the Bayesian statistical term, the 
prior distribution of the transient class, where $k$ represents the class, $\sum_k \pi(k) = 1$, and $\pi(k)\ge 0$ for all $k$. When a new transient is detected, 
a feature vector, which we denote as $\bm{x}$, is provided. Assuming the 
probability distribution $p(\bm{x}|k)$ is known for each class, the posterior 
distribution of the class is computed with the Bayes theorem as in the 
following equation, and the class with the maximum probability is the 
estimated class.
\begin{eqnarray}
p_{\bm{x}}(k)\coloneqq
p(k|\bm{x}) = \frac{p(\bm{x}|k)\pi(k)}{\sum_k p(\bm{x}|k)\pi(k)}.\label{eq:Bayes_th}
\end{eqnarray}
The classification is certain if the probability of one class is much larger 
than that of the others; otherwise, it is not. The uncertainty of the 
classification is well represented by the Shannon entropy:
\begin{eqnarray}
    H_{\bm{x}}(K) = -\sum_k p_{\bm{x}}(k)\log_2 p_{\bm{x}}(k).
\end{eqnarray}
The unit of the Shannon entropy is bits. The entropy is maximized to 
$\log_2 5$ when the class is totally unknown ($p_{\bm{x}}(k) = 1/5$), and 
minimized to $0$ when the class is known ($p_{\bm{x}}(k) = 1$ for one of the 
classes). 

When a transient is detected, the uncertainty regarding its classification is 
typically high because the information obtained from  $\bm{x}$ is limited. In 
order to achieve a certain classification, making an appropriate follow-up 
observation is crucial. The strategy of Smart Kanata is to choose the follow-up observation mode that will minimize the uncertainty.

Let $m$ denote the possible mode of the follow-up observation. For each mode, 
we define a measurement ${y}_m$. This is a virtual measurement because the 
real measurement is obtained after the follow-up observation is executed. 
Again, we assume the probability distribution $q(y_m|k)$ is known for each 
class and each observation mode. 

Suppose a follow-up observation mode $m$ is chosen and a virtual measurement 
$y_m$ is observed, the posterior distribution and the Shannon entropy is 
defined as follows:
\begin{equation}
    q_{\bm{x}}(k|y_m)\coloneqq
    q(k|y_m,\bm{x}) 
    =
    \frac{q_{\bm{x}}(k,y_m)}
    {q_{\bm{x}}(y_m)}
    =
    \frac{q(y_m|k)p_{\bm{x}}(k)}{\sum_k q(y_m|k)p_{\bm{x}}(k)}
\end{equation}
\begin{equation}
    H_{\bm{x}}(K|y_m)=-\sum_k q_{\bm{x}}(k|y_m)\log_2 q_{\bm{x}}(k|y_m).
\end{equation}
Here, we defined $q_{\bm{x}}(k,y_m)\coloneqq q(k,y_m|\bm{x})$ and $q_{\bm{x}}
(y_m)\coloneqq q(y_m|\bm{x})$.

The expected value of $H_{\bm{x}}(K|{y_m})$, known as the conditional 
entropy, is calculated as follows:
\begin{eqnarray}
    H_{\bm{x}}(K|Y_m) &=& \int q_{\bm{x}}(y_m)H_{\bm{x}}(K|y_m)dy_m \nonumber\\ 
    &=& 
    -\int 
    \sum_k q_{\bm{x}}(k,y_m)\
    \log_2 
    \frac{q_{\bm{x}}(k,y_m)}{q_{\bm{x}}(y_m)}
    dy_m.
    \label{eq:cond_entro}
\end{eqnarray}
It is easy to show that $H_{\bm{x}}(K)\ge H_{\bm{x}}(K|Y_m)$ holds in 
general. The difference between $H_{\bm{x}}(K)$ and $H_{\bm{x}}(K|Y_m)$ is 
called the mutual information (MI) in information theory. 
\begin{eqnarray}
    I_{\bm{x}}(K;Y_m) &=& H_{\bm{x}}(K) - H_{\bm{x}}(K|Y_m) \nonumber \\
    &=& 
    \int 
    \sum_k q_{\bm{x}}(k,y_m)\
    \log_2 
    \frac{q_{\bm{x}}(k,y_m)}{q_{\bm{x}}(y_m)p_{\bm{x}}(k)}
    dy_m.\label{eq:mi}
\end{eqnarray}
This is nonnegative and defines the expected gain of the information by 
choosing the observation mode $m$. Smart Kanata chooses the most informative 
observation mode, that is, the mode $m$ which maximizes $I_{\bm{x}}(K;Y_m)$ 
(or equivalently, minimizes $H_{\bm{x}}(K|Y_m)$). 

This framework requires probability models $\pi(k)$, $p(\bm{x}|k)$, and 
$q(y_m|k)$. We explain how we defined them in the next subsections.

\subsection{Variable star type classification}

We develop a generative model (GM) that estimates $p_{\bm{x}}(k)$ by 
modeling $\pi(k)$ and $p(\bm{x}|k)$ in equation~(\ref{eq:Bayes_th}). 

For the prior probability $\pi(k)$ in Equation~(\ref{eq:Bayes_th}), we assume 
a flat distribution: $\pi(k) = (0.2, 0.2, 0.2, 0.2, 0.2)$. Ideally, this 
prior distribution should reflect the distribution of real targets in Smart 
Kanata. However, estimating the distribution based on the population of each 
class of variables is challenging, as the actual samples are generated 
through the complex detection processes of various surveys. The prior 
distribution will be updated as the number of actual samples increases during 
the operation of Smart Kanata.

We approximate $p(\bm{x}|k)$ as multivariate normal distributions, given 
by:
\begin{equation}
p(\bm{x}|k) = \frac{1}{\sqrt{(2\pi)^N|\bm{\Sigma}_k|}}\exp{\left\{-\frac{1}{2}(\bm{x}-\bm{\mu}_k)^T\bm{\Sigma}_k^{-1}(\bm{x}-\bm{\mu}_k)\right\}},\label{eq:multi_normal}
\end{equation}
where $N$ represents the number of features, $\bm{\mu}_k$ is the mean of 
the feature vector, and $\bm{\Sigma}_k$ represents the covariance matrix 
of the features for class $k$, with dimensions $N\times N$. These model 
parameters, $\bm{\mu}_k$ and $\sigma_{ij}$ (the elements of $\bm{\Sigma}$) 
can be estimated as the mean of the sample and the variance/covariance of 
the sample from the training data sets described in the following. 

We collected the training dataset samples from the AAVSO VSX based on the object types as follows:
\begin{description}
    \item[Class N]{Labeled N, NA, NB, NC, and NR.}
    \item[Class DN]{Labeled UGSS, UGZ, UGSU, and UGER.}
    \item[Class WZ]{Labeled UGWZ.}
    \item[Class M]{Labeled M.}
    \item[Class F]{Labeled UV.}
\end{description}
For all classes, we specifically collected objects with recorded 
magnitudes at maximum ($m_{\rm max}$) in the pass-bands $V$, $CV$, $g$, or 
$pg$. We excluded AM CVn stars from our data set because their quiescent 
color and outburst amplitude have distinct distributions from those of 
ordinary dwarf novae \citep{sol10amcvn}. Extragalactic objects were also 
excluded. Additionally, 
to ensure a more evenly distributed dataset, we omitted class M objects 
discovered by OGLE to avoid excessive concentration in the bulge direction 
(\cite{gro05mira, mat05mira}). Furthermore, we only included long-period 
Class M objects with periods exceeding 400 days, as they could be confused 
with CVs. We exclusively extracted bright Class F objects with 
$m_{\rm max} < 16.0$ to obtain magnitude and color information for their 
quiescent counterparts.

We obtain $\ell$, $b$ and $m_{\rm max}$ from the coordinates and the 
magnitude at the time of discovery for new transients. For our training 
dataset, we rely on AAVSO VSX, which provides these features for all the 
samples. We calculate $M_{\rm max}$ if we have access to the distance 
($d$) information from Gaia DR3, without adjusting for interstellar 
extinction \citep{gaiadr3}. To determine the apparent magnitude at 
quiescence, $m_{\rm qui}$, we use the $G$-band mag. from Gaia DR3 whenever 
available. If not, we consider alternative sources in the following order: 
$g$ mag. from Pan-STARRS Release 1 Survey (DR1), $g$ mag. from Sloan 
Digital Sky Surveys (SDSS) Release 16 (DR16), $r$ mag. from Pan-STARRS 
DR1, and $r$ mag. from SDSS DR16 (\cite{panstarrs1, sdssdr16}). $a$ is 
calculated as $m_{\rm qui}-m_{\rm max}$. We obtain $M_{\rm qui}$ when both 
$m_{\rm qui}$ and $d$ are available. For the colors ($g-r$, $r-i$, and 
$i-z$), we source them from Pan-STARRS DR1, and in cases where no 
counterpart is found there, we turn to SDSS DR16. Any color values that 
fall outside the range of $\pm 3\sigma$ compared to the original samples 
for each color are treated as outliers and as missing values in our 
training data set. 
Regarding cross-matching of the catalogs mentioned, we identify the 
counterpart of the VSX sources as the closest object within 
$r_{\rm match}$ in each catalog.

AAVSO VSX undergoes continuous updates. Our training data set was created 
in February 2023.

\begin{figure}
 \begin{center}
  \includegraphics[width=8cm]{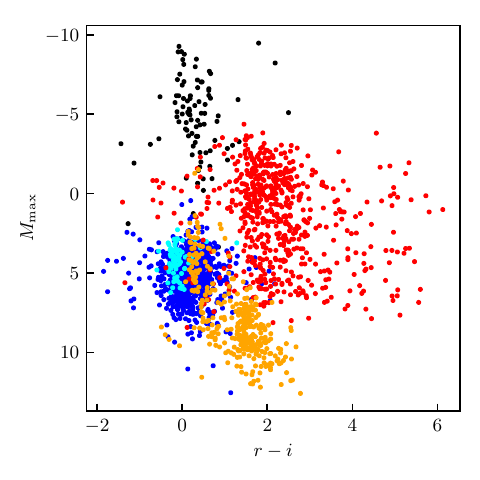} 
 \end{center}
\caption{Scatter plot of the training data set in $M_{\rm max}$ and 
$r-i$. The black, blue, cyan, red, and orange dots represent the samples 
of class N, DN, WZ, M, and F, respectively.
}\label{fig:hr}
\end{figure}

In Figure~\ref{fig:hr}, we provide an example that illustrates the 
distribution of the training data set in the feature space, specifically 
through a scatter plot of $M_{\rm max}$ and $r-i$. The numbers of the 
training dataset, along with the proportion of missing data, are 
summarized in Table~\ref{tab:train}. Our data set comprises a total of 
3686 samples. Notably, a substantial portion of the data table is missing, 
primarily due to the absence of observations during faint quiescent 
phases. This is particularly pronounced in the case of WZ Sge stars, which 
have low luminosity at quiescence, resulting in a high proportion of 
missing data. Novae, on the other hand, exhibit a relatively higher 
luminosity, but their significant distance from Earth results in 
faint quiescent apparent magnitudes, thus contributing to the high proportion 
of missing data.


\begin{table*}
  \tbl{Training data set and fraction of the missing data.}{%
  \begin{tabular}{ccccccccccc}
      \hline
      Types of Var. & Class & N & $\ell$ & $b$ & $M_{\rm max}$ & $M_{\rm qui}$ & $a$ & $g-r$ & $r-i$ & $i-z$\\ 
      \hline
      Nova & N & 332 & 0.00 & 0.00 & 0.53 & 0.53 & 0.13 & 0.50 & 0.49 & 0.46\\
      Dwarf nova & DN & 1528 & 0.00 & 0.00 & 0.30 & 0.30 & 0.05 & 0.23 & 0.23 & 0.25\\
      WZ Sge-type DN & WZ & 232 & 0.00 & 0.00 & 0.61 & 0.62 & 0.29 & 0.42 & 0.44 & 0.53\\
      Mira & M & 1062 & 0.00 & 0.00 & 0.14 & 0.14 & 0.00 & 0.32 & 0.32 & 0.34\\
      Flare star & F & 532 & 0.00 & 0.00 & 0.22 & 0.22 & 0.12 & 0.29 & 0.28 & 0.27\\
      \hline
    \end{tabular}}\label{tab:train}
\end{table*}

The distributions of the training data are approximated as normal 
distributions in equation~(\ref{eq:multi_normal}), which give $p(\bm{x}|k)$. 
Their means and variance-covariance matrices are shown in Appendix 1.  
Figure~\ref{fig:gm_hr} displays the marginal probability distribution of 
$p(\bm{x}|k)$ corresponding to Figure~\ref{fig:hr}, that is, within the 
$M_{\rm max}$--$(r-i)$ plane. 

It should be noted that, in the classification of actual new transients, the 
available features vary from object to object. For example, if only three 
features,$\ell$, $b$, and $M_{\rm max}$, are available due to the absence of
the quiescent counterpart, then $\bm{\mu}$ consists of these three elements
and $\bm{\Sigma}$ contains a matrix of elements $3\times 3$.

\begin{figure}
 \begin{center}
  \includegraphics[width=8cm]{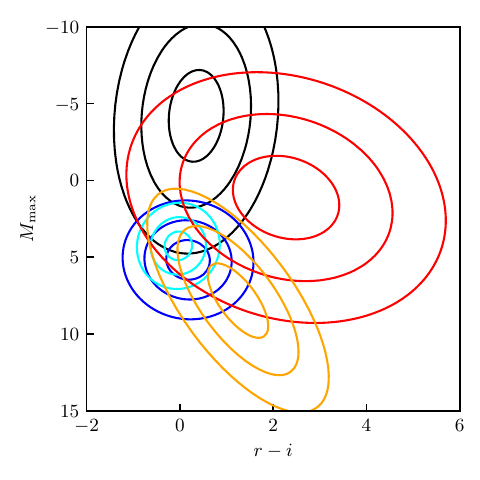} 
 \end{center}
\caption{Marginal probability distributions of $p(\bm{x}|k)$ on the 
$M_{\rm max}$--$(r-i)$ plane. The black, blue, cyan, red, and orange curves 
depict that of class N, DN, WZ, M, and F, respectively. For the observed 
distribution, see Figure~\ref{fig:hr}.
}\label{fig:gm_hr}
\end{figure}

We evaluated the classifier's performance using leave-one-out 
cross-validation (LOOCV) with the training dataset \citep{has17elements}. The LOOCV accuracy was 
determined to be 0.813, and the confusion matrix is shown in 
Table~\ref{tab:conf_all}. These results are based on all 
samples, including those with missing values for all features except $(\ell, 
b)$. There are 239 such samples, and they represent a challenge for accurate 
classification. When these samples are excluded, the classifier performance
improves, achieving an accuracy of 0.869. Table~\ref{tab:conf_limit} is the 
confusion matrix for this improved performance. This table shows a reduction 
in the number of misclassified samples compared to Table~\ref{tab:conf_all}. 
In both cases, a substantial portion of the DN samples is misclassified as 
WZ. This occurs because they occupy overlapping regions within the feature 
space. The information given by the eight features is not sufficient to classify them perfectly. It is essential to note that the primary objective of Smart Kanata 
is not to develop a high-performance classifier but rather to facilitate 
optimal follow-up observations. Consequently, the presence of classes that
cannot be precisely separated is not significant.

\begin{table}
  \tbl{Confusion matrix of the classifier for all samples in the training data set.}{%
  \begin{tabular}{cccccc}
      \hline
      \diagbox{Pred.}{Label} & N & DN & WZ & M & F\\
      \hline
      N  & 292 &  41 &  16 &   0 &  17 \\
      DN &  13 &1070 &  43 &   2 &  35 \\
      WZ &  12 & 320 & 145 &   1 &  16 \\
      M  &  12 &  37 &  25 &1048 &  22 \\
      F  &   3 &  60 &   3 &  11 & 442 \\
      \hline
      Accuracy & 0.880 & 0.700 & 0.625 & 0.987 & 0.831\\
      \hline
    \end{tabular}}\label{tab:conf_all}
\end{table}

\begin{table}
  \tbl{Confusion matrix of the classifier for the samples which have more than three features.}{%
  \begin{tabular}{cccccc}
      \hline
      \diagbox{Pred.}{Label} & N & DN & WZ & M & F\\
      \hline
      N  & 264 &  18 &   3 &   0 &   9 \\
      DN &  13 &1050 &  19 &   2 &  33 \\
      WZ &   7 & 310 & 144 &   0 &   0 \\
      M  &  12 &  17 &   0 &1048 &   9 \\
      F  &   1 &  57 &   0 &  11 & 420 \\
      \hline
      Accuracy & 0.889 & 0.723 & 0.867 & 0.988 & 0.892\\
      \hline
    \end{tabular}}\label{tab:conf_limit}
\end{table}

\subsection{Decision making of the follow-up observation mode}

The follow-up observation modes $B-V$, EW, and Var are indexed as $m=1$, $2$, 
and $3$, respectively. As described in \S2.2, we need $q(y_m|k)$ to compute 
$I_{\bm{x}}(K;Y_m)$ in addition to $\pi(k)$ and $p(\bm{x}|k)$. We established 
$q(y_m|k)$ based on observed data, as follows.

\begin{figure*}
 \begin{center}
  \includegraphics[width=16cm]{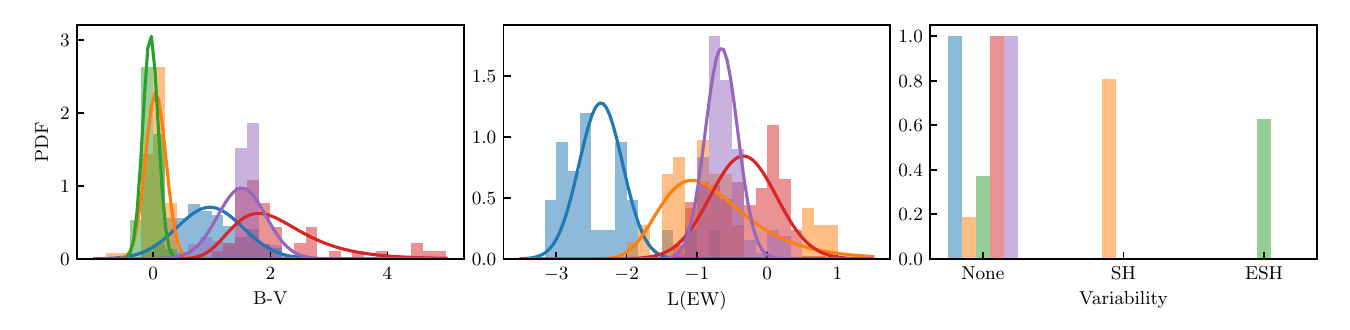} 
 \end{center}
\caption{Observed distributions of $y_m$, that is, $B-V$ ($m=1$, left), EW of 
H$\alpha$ ($m=2$, middle), and short-term variability ($m=3$, right). 
The blue, orange, green, red, and purple curves depict the probability 
density functions for class N, DN, WZ, M, and F, respectively. In the middle 
panel, the orange histogram and curve represent the combined class of classes 
DN and WZ.
}\label{fig:likel}
\end{figure*}

First, we examine the $B-V$ distributions ($m=1$) for four classes: N, DN, 
WZ, and M, drawing data from the AAVSO database. For the samples of our 
training dataset discussed in \S~2.2, the $B-V$ values at their maximum 
brightness were calculated using simultaneous observations in the $B$ and $V$ 
bands obtained from AAVSO. For classes N, DN, and WZ, the data at maximum was 
defined as observations made within a 2-day window from the observed peak. In 
the case of class M, this window was extended to 10 days because their 
variation time-scale is longer than the other classes. For class F, we used 
their quiescent color because flares typically have very short time-scales, 
and our follow-up observations are expected to capture their quiescent state 
following the flare. The $B-V$ values for class F objects were derived from 
their $g-r$ values in Pan-STARRS DR1, applying the transformation described 
in \citet{ton12sdssphot}. As a result, we obtain the observed $B-V$ 
distributions for all five classes, as shown in the left panel of 
Figure~\ref{fig:likel}.

We approximate the observed distributions using two types of probability 
distributions to obtain $q(y_m|k)$. For apparently symmetric distributions, 
we employ normal distributions $\mathcal{N}(\mu, \sigma)$. On the 
other hand, for highly asymmetric distributions, we use Gumbel distributions $\mathcal{G}(\mu, \beta)$ whose density function is defined as:
\begin{eqnarray}
    f(x) &=&  \frac{1}{\beta}\exp\left\{ -\frac{x-\mu}{\beta}-\exp\left( -\frac{x-\mu}{\beta} \right) \right\}
\end{eqnarray}
In both cases, we estimate the parameters $\mu$, $\sigma$, and $\beta$ from 
the available samples. A summary of these $q(y_m|k)$, is shown in 
Table~\ref{tab:likel}.

\begin{table*}
  \tbl{List of $q(y_m|k)$, probability distributions of $y_m$ for each class.}{%
  \begin{tabular}{ccccccc}
      \hline
          & $k=$ & 1 & 2 & 3 & 4 & 5\\
      $m$ & Mode & N & DN & WZ & M & F\\
      \hline
      1 & $B-V$ & $\mathcal{N}(0.97,0.56)$ 
                & $\mathcal{N}(0.04,0.18)$ 
                & $\mathcal{N}(-0.05,0.13)$ 
                & $\mathcal{G}(1.80,0.59)$ 
                & $\mathcal{N}(1.51,0.41)$ \\
      2 & EW    & $\mathcal{N}(-2.37,0.31)$ 
                & $\mathcal{G}(-1.07,0.57)$ 
                & $\mathcal{G}(-1.07,0.57)$
                & $\mathcal{N}(-0.34,0.47)$
                & $\mathcal{N}(-0.65,0.23)$ \\
      3 & Var$^*$ & [1.00, 0.00, 0.00] 
                  & [0.19, 0.81, 0.00] 
                  & [0.20, 0.00, 0.80] 
                  & [1.00, 0.00, 0.00]
                  & [1.00, 0.00, 0.00]\\
      \hline
    \end{tabular}}\label{tab:likel}
    \begin{tabnote}
 \footnotemark[$*$] The three elements represent the detection probabilities of $[$no periodic variation, ordinary superhumps, early superhumps$]$.
  \end{tabnote}
\end{table*}

Second, we examine the distribution of H$\alpha$ EW ($m=2$) for the five
classes. For classes N, DN, and WZ, we gathered EW data from 
Astronomer's Telegram (ATel)\footnote{https://www.astronomerstelegram.org} 
and published papers. Specifically, for class N, we considered EW data 
observed within a 10-day window from the discovery. Most of the data for
class N originated from ATel, while the data for classes DN and WZ primarily 
came from \citet{tan21cvspec} and \citet{han20cvspec}. The dataset sizes for 
classes N, DN, and WZ were relatively small, comprising 26, 28, and 15 
samples, respectively. Given the absence of significant differences in the 
distributions between classes DN and WZ, we combined the data from these two 
classes. For class M, we calculated EW for 257 Mira-type objects, 
using the LAMOST database \citep{yao17mira}. For class F, we relied on 351 
EW measurements from quiescent UV Cet stars reported in \citet{ger99uvcet}.

EW values are positive for the absorption lines and negative for the
emission lines. In the case of class N, most EW values are below 
$-100$\,\AA\ and can even reach as low as $-1000$\,\AA, whereas for other 
classes, they mainly fall between $-1$\,\AA\ and $0$\,\AA. To effectively 
handle the distribution of EW that covers three orders of 
magnitude, we transform EW using the function:
\begin{eqnarray}
    L({\rm EW})&=&{\rm sgn}({\rm EW}_{H\alpha})\log_{10}(1+|{\rm EW}_{H\alpha}|)
\end{eqnarray}
The histograms presented in the middle panel of Figure~\ref{fig:likel} 
illustrate the observed distributions of $L({\rm EW})$. We approximate these 
distributions using the normal and Gumbel distributions as detailed in 
Table~\ref{tab:likel}, which are represented by the curves in the middle 
panel of Figure~\ref{fig:likel}. It is important to note that for class N, we 
estimate the value of $\sigma$ in the observed distribution of $L({\rm EW})$ 
by excluding outliers with exceptionally large $L({\rm EW})$. This adjustment 
enables Smart Kanata to prioritize spectroscopic observations when both class 
N and either class WZ or DN have high class-probabilities.

Third, we employ discrete probability distributions for three types of 
short-term variations which are investigated by mode Var observations 
($m=3$): No periodic variations ($P_{\rm no}$), ordinary superhumps 
($P_{\rm SH}$), and early superhumps ($P_{\rm ESH}$). Ordinary superhumps are 
associated with SU UMa-type dwarf novae and exhibit periodic variations 
typically with a few percent longer period than the binary 
orbital period. In contrast, SS Cyg-type dwarf novae do not display 
superhumps. For class DN, we set $P_{\rm no}$, representing the probability 
that a newly detected event is an outburst of SS Cyg objects, at 0.19, while
$P_{\rm SH}$, indicating the probability for SU UMa objects, is set to 0.81. 
These probabilities are based on the population of dwarf novae identified by 
SDSS \citep{ini23sdssdn}. According to \citet{kat15wz}, the fraction of early 
superhumps detectable with an amplitude greater than 0.02 mag is 0.63. 
Therefore, we set $P_{\rm ESH}=0.63$ and $P_{\rm no}=0.27$ for class WZ. On 
the other hand, classes N, M, and F do not show any periodic variations like 
superhumps or early superhumps, so $P_{\rm no}=1$ for them. The probability 
distributions, represented as $[P_{\rm no}, P_{\rm SH}, P_{\rm ESH}]$, are 
summarized in Table~\ref{tab:likel}. These distributions may include zero 
probabilities, which we replace with small values ($10^{-10}$) for 
calculations in equation~(\ref{eq:cond_entro}).

\subsection{Follow-up observation}

We use two instruments for our follow-up observations: Hiroshima Optical and 
Near-Infrared camera (HONIR; \cite{honir}) and Hiroshima One-shot Wide-field 
Polarimeter (HOWPol; \cite{howpol}). HONIR is mounted at the Cassegrain focus 
of Kanata. It has one CCD and one HgCdTe array, enabling us to simultaneously 
obtain optical and near-infrared images. We use it for the imaging 
observation in Smart Kanata, that is, mode $B-V$ and Var. For events with a 
discovery magnitude of $>12.0$~mag, we set an exposure time of 60~s, while 
for brighter events, it is reduced to 30~s. We collect a set of $B$ and $V$ 
band images for mode $B-V$, while we obtain  time-series $V$ and $J$ band images covering a 
four-hour duration for mode Var. A four-hour run covers 2--3 orbital periods of WZ Sge stars, thereby allowing us to identify the nature of short-term variations. HOWPol is attached to the Nasmyth focus and 
is used for spectroscopic observations. We obtain a low-resolution ($R\sim 300$) spectrum with a wavelength coverage from 4000\AA\, to 8500\AA\, using HOWPol in mode EW. The exposure time for spectroscopy is set at 120~s. 

\begin{figure}
 \begin{center}
  \includegraphics[width=8cm]{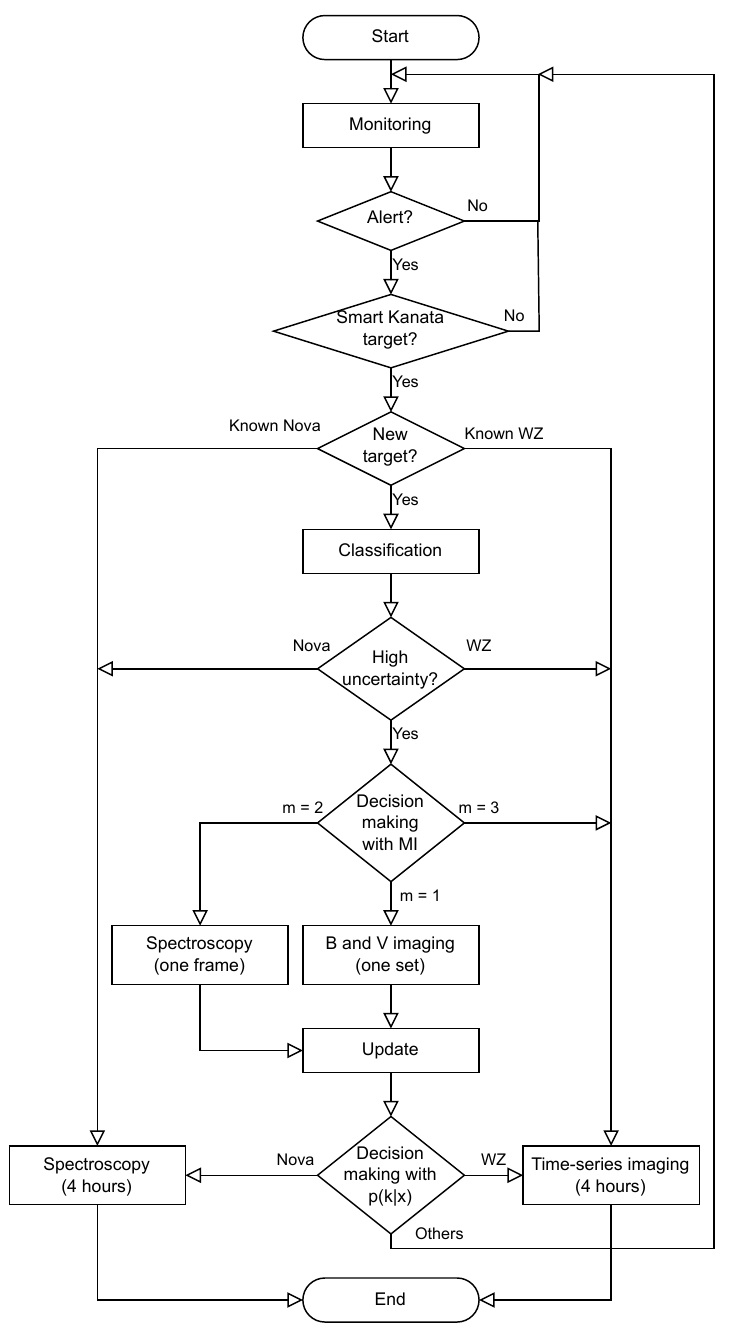} 
 \end{center}
\caption{Flowchart of the follow-up observation process in Smart Kanata.
}\label{fig:flow}
\end{figure}

Figure~\ref{fig:flow} shows the flowchart of the observation process. 
After receiving alerts from online platforms and selecting potential targets 
by excluding SNe and AGN activity as described in Section 2.1, Smart Kanata 
initiates follow-up observations. These observations fall into three 
scenarios.
 
In the first case, Smart Kanata triggers the observation 
when it detects transient events belonging to known novae or WZ Sge-type DN 
(\S~2.1). For novae, it initiates spectroscopic observations, whereas for WZ 
Sge-type objects, it initiates time-series imaging. In both modes, the observations will continue for 4 hours or until dawn. The second 
case involves triggering the observation when the object is classified as 
class N or WZ with a high probability: $p(k|\bm{x})>p_{\rm obs}$. The default 
value of $p_{\rm obs}$ is 0.95, as shown in Table~\ref{tab:smartk_params}. 
The observation modes follow the first case. 

In the third case, Smart Kanata triggers the observation with the highest $I_{\bm{x}}(K;Y_m)$ mode when 
$p_{\rm target}=p({\rm N}|\bm{x})+p({\rm WZ}|\bm{x})>0.1$. 
In this case, after completing the initial follow-up observation process, 
we update the class probability and determine the next observation mode. We 
calculate the updated class probability using the observed value 
$y_{m, {\rm obs}}$ according to the formula:
\begin{eqnarray}
p^\prime(k|y_{m, {\rm obs}},\bm{x}) &=& \frac{q(y_{m, {\rm obs}} | k) p(k|\bm{x})}{\sum_k q(y_{m, {\rm obs}} | k) p(k|\bm{x})}
\end{eqnarray}
If class N has the highest $p^\prime$, we perform time-series spectroscopic 
observations, while if class WZ has the highest probability, we perform 
time-series imaging. In other cases, the program returns to the main loop.

Before triggering any of the mentioned observation processes, we perform 
checks on conditions to confirm whether observation is feasible. These checks
include the following parameters: time, position in the sky, telescope
availability, and weather. The observable time is restricted to
periods before and after civil twilight. The telescope has a hardware 
limitation, setting the lowest observable altitude to 10 degrees. 
Observations are only initiated when the telescope's server program is 
operational. To check weather conditions, we implemented a weather classifier 
using a convolutional neural network (CNN), employing data from our all-sky 
camera located on the observatory roof. Additional details about this weather 
classifier are provided in the Appendix 2.

\section{Results}
\subsection{System evaluation}

Smart Kanata was partially operational from 2021 to 2023 while under 
development. During this phase, only the time-series imaging mode, that is,  
mode Var, was available for follow-up observations. 
On 11 November 2022, the system successfully performed its first automated 
observation, targeting ZTF22abteqxm\footnote{TNS: https://www.wis-tns.org/object/2022zxb}. The object was confirmed to be a WZ Sge-type object ZTF22abteqxm\footnote{[vsnet-alert 27099] 
http://ooruri.kusastro.kyoto-u.ac.jp/mailman3/hyperkitty/list/vsnet-alert@ooruri.kusastro.kyoto-u.ac.jp/message/NULUCDEKJRSR6CRVW5WMWRPUAOJWVULU/}.
The results of this observation will be detailed in a forthcoming paper (Sazaki, et al., in prep.).

Following the addition of multicolor imaging and spectroscopy modes, Smart 
Kanata entered full operational status on 18 December 2023. 
Table~\ref{tab:target} lists the events that Smart Kanata attempted to 
initiate following-up observations during this phase, covering
the period until October 2024. 
The table does not include targets observed during the partial operation 
phase mentioned above.

Out of the 21 opportunities for automated
follow-up observations, the system completed two observations successfully. 
In three cases, the system stopped follow-up observations as the target 
probability, $p_{\rm target}< 0.1$. The remaining 16 observation
opportunities were missed due to poor weather (nine cases) or the 
observatory being closed (six cases) or system failure (one case).

In Table~\ref{tab:target}, the maximum values for class probabilities are 
highlighted in bold. The confusion matrix for the classifier is shown in 
Table~\ref{tab:conf_real}. We defined the prediction of the classifier based 
on the class with the highest probability among the five possible classes to 
produce this confusion matrix. For two cases where the dwarf nova subclass 
was ambiguous, we assigned a value of 0.5 to the DN and WZ classes in 
Table~\ref{tab:conf_real}. The classifier predictions matched the true 
classes in 11.5 cases, yielding an accuracy of 0.575. Among the 8.5 
misclassified objects, three cases were objects that the classifier did not 
account for (two microlensing events and one GRB afterglow), and 5.5 cases 
had high uncertainties, with the highest probability in each case less than 
0.5.

In three cases, MI was less than 0.001, indicating low class 
uncertainty. For the other 18 cases, MI was highest for mode $B-V$ in five 
instances, mode EW in five, and mode Var in eight. The system tended to 
select the mode $B-V$ when the uncertainty of the class was high, while mode 
EW was most likely chosen when there was a significant probability of the 
class N. Mode Var was typically selected when the probability of class N was 
lower. Overall, the selections of the follow-up mode were appropriate and 
there were no clearly incorrect decisions.

We successfully conducted automated observations of TCP J17395720$-$2627410 
(V4370 Oph) and TCP J17440698$-$2125195. Details of these objects are 
discussed in the following subsection. Notes on other individual objects are 
provided below:

\begin{landscape}
\begin{table}
\tbl{Class probability and MI for our targets.}{%
  \begin{tabular}{ccccccc}
  \hline
  Time (UT)& ID & Observer & Class probability & MI & Class & Status \\
  & & & [N, DN, WZ, M, F] & [$B-V$, EW, Var.] & &\\
  \hline
  23 Dec. 2023 & ATLAS23xus & ATLAS & $[0.000, 0.167, {\bf 0.833}, 0.000, 0.000]$ & $[0.054, 0.000, {\bf 0.524}]$ & WZ:$^1$ & (Bad weather) \\
  29 Dec. 2023 & Gaia23duw & Gaia Alert& $[0.000, {\bf 0.994}, 0.006, 0.000, 0.000]$ & $[0.002, 0.000, {\bf 0.044}]$ & DN$^2$ & (Bad weather) \\ 
  31 Dec. 2023 & ATLAS23yai & ATLAS & $[0.000, 0.256, {\bf 0.744}, 0.000, 0.000]$ & $[0.072, 0.000, {\bf 0.662}]$ & WZ: & (Closed) \\
  17 Jan. 2024 & TCP J03205377$+$4227578 & XOSS & $[0.000, 0.000, 0.000, 0.000, {\bf 1.000}]$ & $[0.000, 0.000, 0.000]$ & F$^3$ & ($p_{\rm target}<0.1$)\\
  23 Jan. 2024 & TCP J07593036$+$0459025 & Nakamura, Y. & $[0.003, 0.027, {\bf 0.970}, 0.000, 0.000]$ & $[0.037, 0.024, {\bf 0.151}]$ & DN$^4$ & (Bad weather)\\
  25 Jan. 2024 & MASTER OT J174112.75$+$232535.1 & MASTER & $[0.011, 0.018, {\bf 0.971}, 0.000, 0.000]$ & $[0.079, 0.070, {\bf 0.130}]$ & WZ: & (Bad weather)\\ 
  10 Mar. 2024 & TCP J17395720$-$2627410/V4370 Oph & Kojima, T. & $[{\bf 0.997}, 0.000,  0.000, 0.000, 0.002]$ & $[0.003, {\bf 0.027}, 0.000]$ & N & Success\\
  19 Mar. 2024 & ASASSN-24ca & ASAS-SN & $[{\bf 0.506}, 0.257, 0.189,  0.004, 0.045]$ & $[0.778, 0.940, {\bf 0.978}]$ & N$^5$ & (Closed)\\
  26 Mar. 2024 & TCP J18595064$+$0605336 & Ueda, S. & $[{\bf 0.452}, 0.122, 0.092, 0.257, 0.078]$ & $[0.903, {\bf 1.062}, 0.680]$ & F & (Closed)\\
  12 May 2024 & TCP J14231331$-$3658253 & BraTS & $[0.104, {\bf 0.273}, 0.248, 0.177, 0.198]$ & $[{\bf 1.119}, 0.666, 1.029]$ & DN:WZ: & (Bad Weather)\\
  15 Jul. 2024 & KATS24N014 & XOSS & $[0.229, 0.211, 0.164, {\bf 0.258}, 0.137]$ & $[{\bf 1.097}, 0.914, 0.916]$ & DN:WZ: & (Bad Weather)\\
  15 Jul. 2024 & KATS24N015 & XOSS & $[0.139, {\bf 0.265}, 0.223, 0.199, 0.174]$ & $[{\bf 1.122}, 0.748, 1.010]$ & WZ: & (Bad Weather)\\
  20 Jul. 2024 & TCP J17490276$-$2324066 & Itagaki, K. & $[{\bf 0.996}, 0.003, 0.001, 0.000, 0.000]$ & $[0.012, {\bf 0.031}, 0.030]$ & N$^6$ & (Closed)\\
  22 Jul. 2024 & TCP J18111965$-$2822416 & Kojima, T. & $[0.000, {\bf 1.000}, 0.000, 0.000, 0.000]$ & $[0.000, 0.000, 0.000]$ & Microlensing$^7$ & ($p_{\rm target}<0.1$) \\
  11 Aug. 2024 & TCP J17440698$-$2125195 & Kojima, T. & $[{\bf 0.771}, 0.000, 0.000, 0.229, 0.000]$ & $[0.342, {\bf 0.762}, 0.000]$ & Microlensing & Success \\
  9 Sep. 2024 & TCP J17064645$-$3528079 & Itagaki, K. & $[{\bf 0.474}, 0.115, 0.090, 0.249, 0.073]$ & $[0.881, {\bf 1.061}, 0.663]$ & N$^8$ & (Closed)\\
  17 Sep. 2024 & TCP J22481420$+$0629564 & Nakamura, Y. & $[0.000, 0.265, 0.338, 0.011, {\bf 0.386}]$ & $[1.015, 0.277, {\bf 1.050}]$ & -- & (Closed)\\
  20 Sep. 2024 & PSP24V & XOSS & $[0.000, 0.000, 0.000, 0.000, {\bf 1.000}]$ & $[0.000, 0.000, 0.000]$ & F$^9$ & ($p_{\rm target}<0.1$)\\
  4 Oct. 2024 & MASTER OT J065054.42$+$593625.5 & MASTER & $[0.003, 0.025, {\bf 0.972}, 0.000, 0.000]$ & $[0.033, 0.020, {\bf 0.142}]$ & WZ$^{10,11}$ & (Bad Weather) \\
  21 Oct. 2024 & TCP J20233423$+$1142122 & Itagaki, K. & $[{\bf 0.291}, 0.178, 0.141, 0.260, 0.131]$ & $[{\bf 1.053}, 0.993, 0.849]$ & F$^{12}$ & Failure \\
  29 Oct. 2024 & MASTER OT J214120.16$+$050453.4 & MASTER & $[0.014, 0.299, 0.304, 0.073, {\bf 0.310}]$ & $[{\bf 1.061}, 0.355, 1.055]$ & GRB$^{13}$ & (Bad weather)\\
  \hline
  \end{tabular}}\label{tab:target}
  \begin{tabnote}
  \footnotemark[$^1$]{vstnet-alert 27977}
  \footnotemark[$^2$]{\citet{hod23gaia23duw}}
  \footnotemark[$^3$]{http://xjltp.china-vo.org/psp24d.html}
  \footnotemark[$^4$]{vsnet-alert 27986}
  \footnotemark[$^5$]{\citet{ATel16550}}
  \footnotemark[$^6$]{\citet{ATel16727}}
  \footnotemark[$^7$]{http://www.cbat.eps.harvard.edu/unconf/followups/J18111965-2822416.html}
  \footnotemark[$^8$]{\citet{ATel16808}}
  \footnotemark[$^9$]{http://xjltp.china-vo.org/psp24v.html}
  \footnotemark[$^{10}$]{\citet{ATel16858}}
  \footnotemark[$^{11}$]{vsnet-alert 28024}
  \footnotemark[$^{12}$]{http://www.cbat.eps.harvard.edu/unconf/followups/J20233423+1142122.html}
  \footnotemark[$^{13}$]{https://www.wis-tns.org/object/2024zse}
  \end{tabnote}
\end{table}
\end{landscape}

\begin{table}
  \tbl{Confusion matrix of the classifier for the real data set.}{%
  \begin{tabular}{ccccccc}
      \hline
      \diagbox{Pred.}{Label} & N & DN & WZ & M & F & Others\\
      \hline
      N  & 4 & 0 & 0 & 0 & 2 & 1\\
      DN & 0 & 1.5 & 1.5 & 0 & 0 & 1\\
      WZ & 0 & 1 & 4 & 0 & 0 & 0\\
      M  & 0 & 0.5 & 0.5 & 0 & 0 & 0\\
      F  & 0 & 0 & 0 & 0 & 2 & 1\\
      \hline
      Accuracy & 1.000 & 0.500 & 0.667 & --- & 0.500 & ---\\
      \hline
    \end{tabular}}\label{tab:conf_real}
\end{table}

\noindent
{\bf ATLAS23yai}: \citet{zha24ATLAS23yai} identified this object as a dwarf 
nova based on optical spectra. ASAS-SN observations indicated a peak 
brightness of 14.3 mag, with an outburst duration of more than eight days, 
followed by rebrightening events after a temporary fade. The activity 
persisted for over 32 days from detection to the end of the rebrightening 
events. Its quiescent counterpart was found in the Pan-STARRS catalog with 
$g=21.85$, suggesting an outburst amplitude of $\sim 7.5$ mag. This large 
amplitude and prolonged activity suggest that the object is likely of class 
WZ rather than DN.

\noindent
{\bf MASTER OT J174112.75$+$232535.1}: ASAS-SN observations recorded the object showing an exponential decay at a rate of 
$\sim 0.1\;{\rm mag}\,{\rm d}^{-1}$ over ten days. Following this, the object faded below the detection limit, consistent 
with a superoutburst of a dwarf nova. The discovery 
report\footnote{https://www.wis-tns.org/object/2024aud} noted an outburst 
amplitude of 8.1 mag. Despite the absence of spectral data, the light curve 
and high amplitude strongly favor classification as a WZ-type object. 

\noindent
{\bf TCP J18595064$+$0605336}: This flare star was misclassified by Smart 
Kanata. The system assigned the highest probability of 0.452 to class N, 
followed by 0.257 to class M, and a smaller probability of 0.078 to class F. 
This misclassification was caused by a large positional error in the 
discovery report. No quiescent counterpart was found within 
$r_{\rm match}=2\arcsec.0$, though a candidate YSO located $2\arcsec.2$ from 
the reported position ($g=16.8$) was identified \citep{zar18yso}. Combined 
with ROSAT X-ray data, the object is classified as class F \citep{fre22xstar}.

\noindent
{\bf TCP J14231331$-$3658253}: Discovered by Brazilian Transient Search (BraTS)\footnote{http://sites.mpc.com.br/holvorcem/obs/SkySift\_disc\_transients.html} at 15.1 mag, this object 
exhibited a steady fading of $\sim 0.2\;{\rm mag}\,{\rm d}^{-1}$ over six 
days, consistent with either class DN or WZ. No quiescent counterpart was 
found in available catalogs, implying a large outburst amplitude. However, 
due to the short focal length of the BraTS system (616 mm), there may be a 
significant positional error. A Gaia object with $G=18.4$ lies $4\arcsec.3$ 
away, and if this is the quiescent counterpart, the outburst amplitude would 
be smaller, $\sim 3$ mag. Thus, the subclass of this dwarf nova remains 
uncertain.

\noindent
{\bf KATS24N014}: A spectrum reported by \citet{zha24kats24n014} showed a 
blue continuum with a narrow H$\alpha$ emission line, suggesting the object 
belongs to class DN or WZ. ASAS-SN observations confirmed an outburst lasting 
at least five days. The absence of a quiescent counterpart implies a large 
outburst amplitude, but the precise subclass remains unclear.

\noindent
{\bf KATS24N015}: \citet{zha24kats24n014} report a spectrum showing a blue 
continuum with Balmer absorption lines, indicating that the object likely 
belongs to class DN or WZ. The ASAS-SN light curve reveals a main outburst 
lasting more than 12 days, followed by a temporary fading and subsequent 
rebrightenings. The object’s active phase has persisted for over 40 days. Its 
quiescent counterpart, with $b=21.43$, was identified in the APM-North 
Catalogue \citep{APM-North-Cat}, suggesting an outburst amplitude of 
$\gtrsim 7$\,mag. The extended activity and large amplitude strongly support 
the classification of the object as a member of class WZ.

\noindent
{\bf TCP J20233423$+$1142122}: 
Discovered by Itagaki at 10.2 mag, this object exhibited a rapid decline to 
13.2 mag within 25 min.\footnote{http://www.cbat.eps.harvard.edu/unconf/followups/J20233423+1142122.html}. Smart Kanata did not identify a quiescent counterpart, assigning 
the highest probability of 0.291 to class N. Although the system tried to 
initiate an automated observation in mode $B-V$, the observation was canceled 
due to a network connection error. A significant positional discrepancy 
exists in the discovery report, with a Gaia DR3 source of $G=14.08$ located 
$3\arcsec.59$ from the reported position. Based on this and ROSAT X-ray data, 
the object is classified as class F.

\subsection{Initial results}
\subsubsection{Classical nova: V4370 Oph}
A new object with 11.5~mag was discovered by T. Kojima in an image taken on
10.784 (UT) March 2024 and was promptly reported to TOCP. It was designated 
as TCP J17395720$-$2627410. No object brighter than 15~mag was detected at 
the corresponding position in an image taken on 9.81 (UT) March 2024 by M. 
Yamamoto.\footnote{http://www.cbat.eps.harvard.edu/unconf/followups/J17395720-
2627410.html} This discovery report was detected by Smart Kanata, which 
classified the object as a nova with a probability of 0.997 and initiated 
spectroscopic observations using HOWPol automatically. A preliminary analysis 
of the collected data revealed broad Balmer and He\,\emissiontype{I} emission 
lines exhibiting P-Cygni profiles, thus confirming its nova classification 
\citep{uem24atel16521}. Subsequent to this spectroscopic observation, the 
object was assigned the GCVS name V4370 Oph \citep{cbet5365}.

\begin{table}
\tbl{Observation log of V4370 Oph.}{%
  \begin{tabular}{cccc}
  \hline
  MJD$^*$ & $T_{\rm exp}^\dag$ (s) & $\Delta t^\ddag$ (d) & $R^\S$\\
  \hline
  60379.84682 & 120 & 0.06282 & 260\\
  60379.84843 & 120 & 0.06443 & 300\\
  60379.85004 & 120 & 0.06604 & 300\\
  60379.85362 & 120 & 0.06962 & 300\\
  60379.85522 & 120 & 0.07122 & 300\\
  60379.85682 & 120 & 0.07282 & 300\\
  \hline
  60384.84319 & 200 & 5.05919 & 230\\
  60384.84688 & 200 & 5.06288 & 190\\
  60384.85036 & 200 & 5.06636 & 200\\
  \hline
  \end{tabular}}\label{tab:log_v4370}
  \begin{tabnote}
  \footnotemark[$*$]{Time of the exposure center.}
  \footnotemark[$\dag$]{Exposure time.}
  \footnotemark[$\ddag$]{Time since the discovery (2024 March 10.784 UT)}
  \footnotemark[$\S$]{Spectral resolution: $\lambda/\Delta\lambda$}
  \end{tabnote}
\end{table}

Our observation log for V4370 Oph is presented in Table~\ref{tab:log_v4370}. 
The initial exposure started 1.49 hours after discovery. Observations 
persisted for 16.4~min, yielding six frames. After bias subtraction and
flat-field correction, one-dimensional spectra were extracted from the
images. Wavelength calibration was performed using the night-sky lines. 
Flux calibration was omitted because spectrophotometric standard stars were
not observed. In the present study, we exclusively examine spectra
normalized to the continuum flux level. 

\begin{figure*}
 \begin{center}
  \includegraphics[width=17cm]{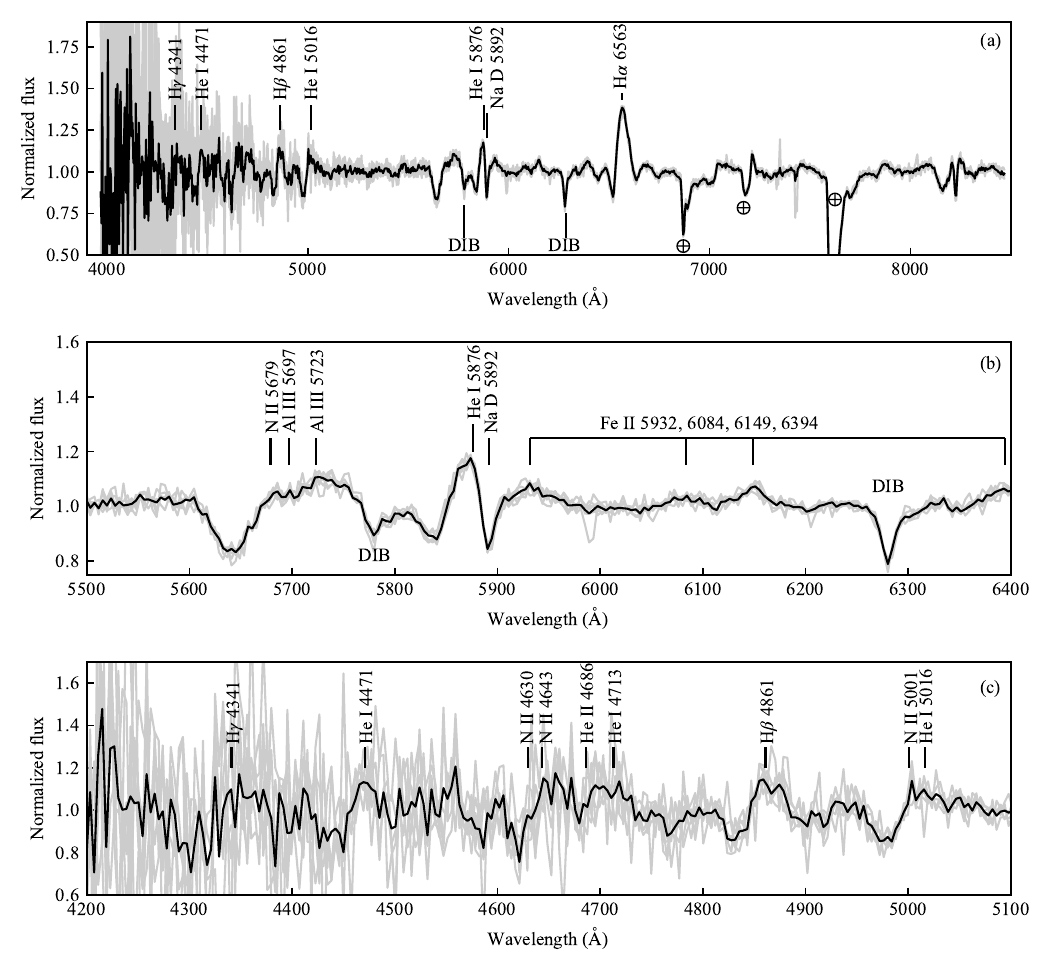} 
 \end{center}
\caption{Spectra of V4370 Oph on 15 March 15 2024. They are normalized 
to the continuum level. Panel (a) shows the entire spectra. Panels (b) and  
(c) show enlarged views in the ranges 5500--6400\,\AA, and 4200--5100\,\AA, 
respectively. The gray lines represent the six spectra, while the black line 
depicts their mean.
}\label{fig:sp_all}
\end{figure*}

Figure~\ref{fig:sp_all} shows the spectra from the first day of
observation. The mean normalized spectrum of the six spectra is represented 
by the black line, while the gray lines illustrate each individual spectrum 
to visualize variances around the mean. Panel~(a) presents the spectra 
throughout the wavelength range, where the Balmer and He\,\emissiontype{I} 
emission lines exhibit prominent P-Cygni profiles. The EW, the full width at 
half maximum (FWHM), and the velocity of the minimum of the absorption 
component ($v_{\rm abs}$) were measured and summarized in Table~\ref{tab:ew}. 
These measurements were conducted for each of the six spectra, with the table 
listing the means and standard deviations of these measurements. The EWs are 
relatively small, approximately $\sim 20\,$\AA, even for H$\alpha$. This 
observation, coupled with the pronounced P-Cygni profiles, suggests that the 
object was in the early stages of a nova eruption.

\begin{table*}
\tbl{Parameters of lines.}{%
  \begin{tabular}{cccccc}
  \hline
  Element & \multicolumn{3}{c}{March 10} & \multicolumn{2}{c}{March 15}\\
          & EW (\AA) & FWHM (km\,s$^{-1}$) & $v_{\rm abs}$ (km\,s$^{-1}$)  & EW (\AA) & FWHM (km\,s$^{-1}$)\\
  \hline
  H$\alpha$ & $20.22\pm 0.68$ & $2531\pm 132$ & $-1999\pm 111$ & $3128\pm 294$ & $6279\pm 90$ \\
  H$\beta$ & $4.35\pm 1.79$ & $1364\pm 432$ & $-1875\pm 165$ & $289\pm 68$ & $5779\pm 91$ \\
  He\,\emissiontype{I} 5876 & $3.46\pm 0.27$ & $1049\pm 420$ & $-1926\pm 156$ & $148\pm 2$ & $6365\pm 99$ \\
  N\,\emissiontype{II} 5679 & --- & --- & --- & $93\pm 8$ & $7596\pm 203$\\
  He\,\emissiontype{I} 7065 & --- & --- & --- & $75\pm 23$ & $5862\pm 85$\\
  O\,\emissiontype{I} 7774 & --- & --- & --- & $143\pm 6$ & $5972\pm 77$\\
  \hline
  \end{tabular}}\label{tab:ew}
\end{table*}

Panels~(b) and (c) provide enlarged views of panel~(a). In panel (b), a 
pronounced Na\,D absorption line is evident. The emission component of 
He\,\emissiontype{I} 5876 likely overlaps with this absorption line. 
Furthermore, an absorption feature around 5650\,\AA\;is noteworthy, likely 
representing a part of a P-Cygni profile, with its emission component
observable in the range 5670--5770\,\AA. This line is tentatively identified 
as N\,\emissiontype{II} 5679, although the width of the emission component
exceeds what can only be attributed to this line. This excess indicates the 
contributions of Al\,\emissiontype{III} 5697 and 5723 to this broad emission
component.

In panel~(c), the P-Cygni profiles of H$\beta$ and N\,\emissiontype{II} 5001 
$+$ He\,\emissiontype{I} 5016 are clearly observed. However, on the bluer 
side, no lines are clearly discernible due to substantial interstellar 
extinction, with $A_V=4.37$ along the line of sight to this object 
\citep{sch11dust}. Nevertheless, indications of H$\gamma$ and 
He\,\emissiontype{I} 4471 are visible, potentially exhibiting P-Cygni 
profiles. Another notable feature is a potential bump in the wavelength range 
4640--4720\,\AA. This bump is consistently observed not only in the mean spectrum, but also in each individual spectrum, supporting its validity. This 
wavelength region encompasses a band rich in high-excitation lines, known as 
the Bowen blend. However, the lack of a strong emission line of 
He\,\emissiontype{II} 4686 suggests that lower excitation lines contribute to 
the emission component. The bump is possibly identified as N\,\emissiontype{II} 
4643, while N\,\emissiontype{II} 4630 is absent. Other candidates include 
moderately excited lines such as N\,\emissiontype{III} 4640 or 
C\,\emissiontype{III} 4650. The contribution of He\,\emissiontype{I} 4713 is 
also plausible.

We analyze short-term variations of the emission lines in our six spectra
collected over a span of 14~min. using a Student $t$ test. Let $f_{ij}$ 
($i=1,2,\cdots,1099$, $j=1,2,\cdots,6$) represent the normalized flux at the 
$i$-th wavelength of the $j$-th spectrum. We modeled temporal variations at 
each wavelength as follows:
\begin{eqnarray}
f_{ij} &=& a_i t_j + b_i + \varepsilon_i\\
\varepsilon_i & \sim & \mathcal{N}(0, \sigma_i^2),
\end{eqnarray}
where $t_j$ denotes the time of the $j$-th spectrum, and $\varepsilon_i$ 
represents Gaussian noise with a variance $\sigma_i^2$. We estimated $a_i$, 
$b_i$, and $\sigma_i^2$ using linear regression for each wavelength and 
tested $a_i\neq 0$ using the Student $t$ distribution.

\begin{figure}
 \begin{center}
  \includegraphics[width=8cm]{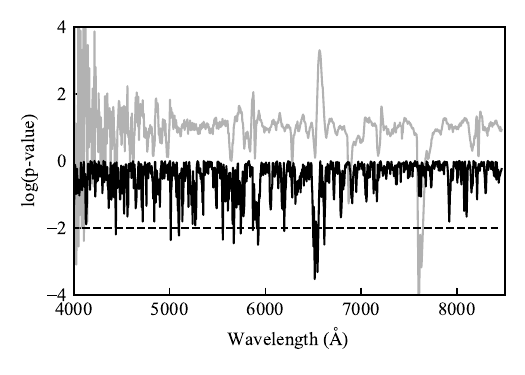} 
 \end{center}
\caption{$p$-values for each wavelength (black line). For details about the 
statistical test, refer to the main text. The gray line represents the mean 
spectrum shown in Figure~\ref{fig:sp_all}. The dashed horizontal line 
represents $p=0.01$.
}\label{fig:pvalues}
\end{figure}

Figure~\ref{fig:pvalues} illustrates the $p$-values for each wavelength. We 
can reject the null hypothesis of $a=0$, indicating that $f_i$ has no 
significant variation, with greater confidence at lower $p$-values. The 
dashed horizontal line represents a significance level of 0.01. The wavelengths that fall sporadically below $p=0.01$ can be considered false
positives. However, a distinct band is observable containing multiple 
consecutive wavelengths with $p<0.01$ centered around H$\alpha$. This 
indicates that a significant variation is present in this wavelength band. 

\begin{figure}
 \begin{center}
  \includegraphics[width=8cm]{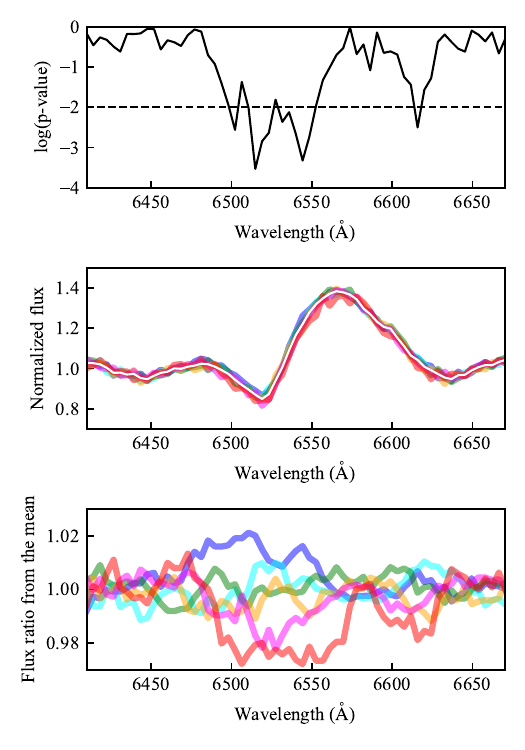} 
 \end{center}
\caption{Top: $p$-values around H$\alpha$. Middle: Spectra around 
H$\alpha$. The colored lines represent the six spectra. Colors progress from 
early (blue) to late (red), with intermediate colors including cyan, green, 
yellow, and magenta. The white line indicates their mean. Bottom: the 
ratio of each spectrum to the mean. To enhance clarity, we display 5-point 
running mean spectra in this panel.
}\label{fig:Halpha_var}
\end{figure}

Figure~\ref{fig:Halpha_var} shows the $p$-values (top) and spectra (middle) 
centered on H$\alpha$. Small $p$-values are notably associated 
with the absorption component and the blue side of the emission component. To 
emphasize potential variations, we present the ratio of each spectrum to the 
mean spectrum in the bottom panel. Consistent with the $p$-values, 
significant deviations from the mean are observed across the spectra, 
particularly from the absorption to the blue side of the emission components. 
This suggests a rapid deepening of the absorption component during our 
observation period.

\begin{figure*}
 \begin{center}
  \includegraphics[width=17cm]{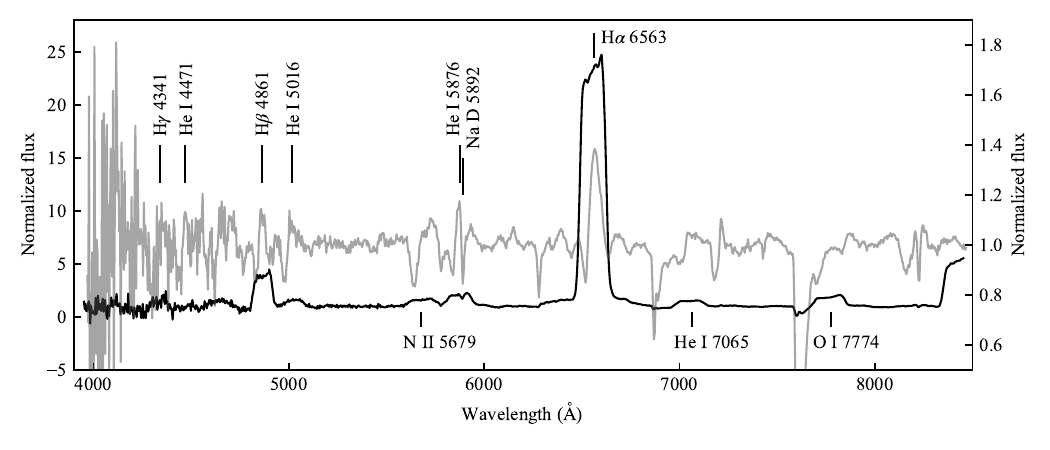} 
 \end{center}
\caption{Spectrum of V4370 Oph on March 15, five days after the discovery 
(black line). The spectrum on March 10 is depicted by the gray 
line.
}\label{fig:sp_max}
\end{figure*}

In addition to the initial observation, we acquired
three spectra on 15 March, five days after the discovery. The mean 
spectrum is depicted in Figure~\ref{fig:sp_max} (black line), with the mean 
spectrum from the first day serving as a reference (gray line). The spectrum 
exhibits typical characteristics for He/N novae after maximum: prominent 
Balmer, He\,\emissiontype{I}, and N\,\emissiontype{II} emission lines without 
any signs of high-excitation lines. The emission lines display a box-like 
profile with a more pronounced red peak. As detailed in Table~\ref{tab:ew}, 
the object underwent a significant increase in both EW and FWHM of the lines 
from the first to the fifth day. These variations in line features
suggest that the high-velocity components grew between the first and fifth 
days.

\subsubsection{Microlensing event: TCP J17440698$-$2125195}

\begin{figure}
 \begin{center}
  \includegraphics[width=8cm]{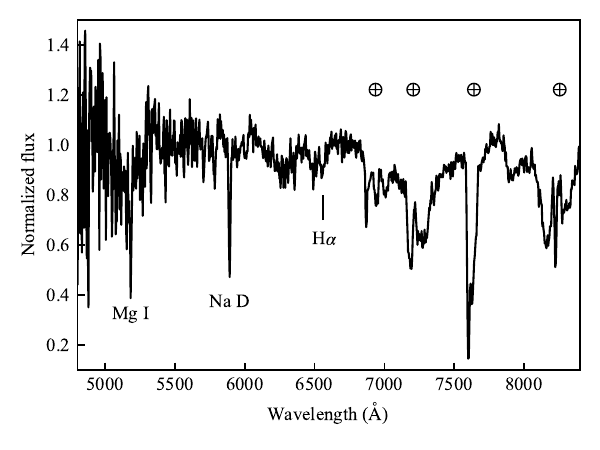} 
 \end{center}
\caption{Spectrum of TCP J17440698$-$2125195 on August 11.
}\label{fig:sp_j1744}
\end{figure}

TCP J17440698$-$2125195 was discovered by T. Kojima at 13.7 mag. on 
11.4556 (UT) August 
2024.\footnote{http://www.cbat.eps.harvard.edu/unconf/followups/J17440698-
2125195.html} Smart Kanata found this discovery from the TOCP report, 
estimated class probabilities of 0.771 for class N and 0.229 for class M, 
made the decision to conduct follow-up observations in mode EW, and started 
spectroscopic observation on 11.53699 (UT) (MJD 60533.53699), $\sim 1.9$ 
hours after the initial discovery. The spectrum is shown in 
Figure~\ref{fig:sp_j1744}. It revealed no prominent emission lines, with 
H$\alpha$ marginally detected in absorption. The EW of H$\alpha$ was measured 
at $\sim 4$\,\AA, prompting the system to update the class probabilities to 
0.000 for class N and 1.000 for class M. Consequently, the system classified 
the object as a Mira and discontinued follow-up observations.

The spectrum shown in Figure~\ref{fig:sp_j1744} is characterized by
strong absorption lines of Na\,D and Mg\,\emissiontype{I}. The presence of 
the Na\,D line is consistent with a significant interstellar extinction in 
the direction of the object ($A_V \sim 2$; \cite{sch11dust}). Along with the 
weak H$\alpha$ absorption line, the Mg\,\emissiontype{I} absorption indicates 
a K-type star. \citet{mun24atel16769} reported spectra taken on 12.863 (UT), 
1.3 days after our observation, when the object had faded by $\sim 
1.3$\, mag. Their spectrum shared common features with ours, and they 
suggested a spectral type of K2 to K3 and a luminosity class of III to II, 
consistent with the quiescent counterpart, which is likely a K-type giant, 
inferred from $V=16.80$ and $B-V=1.7$ ($E(B-V)\sim 0.7$; \cite{NOMAD}).

\begin{figure}
 \begin{center}
  \includegraphics[width=8cm]{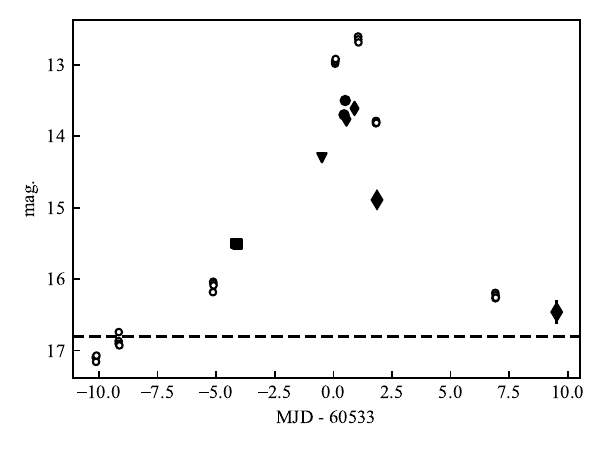} 
 \end{center}
\caption{Light curve of TCP J17440698$-$2125195. The filled squares, triangle, circles, and filled diamonds represents $G$-mag observed by Gaia, visual upper-limit reported in TOCP, $V$-band magnitudes in TOCP  and \citep{mun24atel16769}, and our observation, respectively. The open circles represents the 'orange' band magnitudes observed with ATLAS \citep{ATLAS, ATLAS_transient}. The dashed line indicates the quiescent level. 
}\label{fig:lc_j1744}
\end{figure}

We conducted a follow-up imaging observation and obtained $V=16.46\pm0.17$ on 
20.52148 August (MJD 60542.52148). Photometry was performed using a 
comparison star located at RA$=$17:44:05.15, Dec$=-$21:24:42.8, with 
$V=14.90$ in the NOMAD catalog (\cite{NOMAD}). The light curve of 
TCP J17440698$-$2125195 is shown in Figure~\ref{fig:lc_j1744}. ATLAS and the Gaia 
satellite detected a brightening of the object on 5 and 6 August, respectively. The event was designated as 
Gaia24cfb.\footnote{http://gsaweb.ast.cam.ac.uk/alerts/alert/Gaia24cfb/} These data indicate that the object had already been active six days before
T. Kojima’s discovery. The timescale of variation near the maximum is short: 
$\gtrsim 0.8$\,mag in 1.0 days for the rise and 1.3 mag in 1.0 days for the 
decay. Our observation suggests that the object had nearly reached or was 
still slightly brighter than its quiescent level nine days after its peak 
brightness.

Although Smart Kanata initially classified the object as a Mira, the rapid 
rise and decline around the maximum are inconsistent with this 
classification. In addition to the lack of significant spectral changes, the
gradual rise and decay over approximately 10 days, combined with the rapid
variation around the peak, suggest that this is actually a microlensing event.

Although microlensing events are not included in the list of variable star
classes in Smart Kanata, the spectrum obtained by our system played a
critical role in identifying its nature. This case demonstrates that Smart 
Kanata has the potential for serendipitous discoveries, even for unexpected 
and rare phenomena.

\section{Discussion}

As demonstrated in the previous section, Smart Kanata effectively supports decision making about follow-up observations of CVs. 
However, the GM classifier employed in Smart Kanata provides only a simple, 
quadratic decision boundary, despite the availability of alternative models
that can offer nonlinear decision boundaries. In Section 4.1, we will
compare the performance of other classification models with the GM
classifier. Following this, Section 4.2 will focus on the implications of our 
observations of V4370 Oph.

\subsection{Comparison with the performance of the other classification models}

Logistic regression (LR) provides a linear classification model. The 
probability of multinomial LR is expressed as:
\begin{eqnarray}
    p(\bm{y}|\bm{x},\bm{w}) &=& \frac{\exp(-\bm{w}_k^T\bm{x})}{\sum_j \exp(-\bm{w}_j^T\bm{x})},\label{eq:lr} 
\end{eqnarray}
where $\bm{w}_k$ is the coefficients to be optimized for class $k$, which 
include the bias factor. The objective variable is a "1-of-$K$" encoding 
vector, $\bm{y}=(y_1, y_2, \cdots, y_K)$ such that $y_k=1$ if the sample 
belongs to the class $k$, and $y_i=0$ otherwise. The likelihood and log-
likelihood functions are given as:
\begin{eqnarray}
    \ell(\bm{w}) &=& \prod_i \prod_k \left\{ \frac{\exp(-\bm{w}_k^T\bm{x}_i)}{\sum_j \exp(-\bm{w}_j^T\bm{x}_i)} \right\}^{y_{ik}}\\
    \log \ell(\bm{w}) &=& \sum_i \sum_k y_{ik}\log \frac{\exp(-\bm{w}_k^T\bm{x}_i)}{\sum_j \exp(-\bm{w}_j^T\bm{x}_i)}
\end{eqnarray}

A nonlinear LR classifier can be obtained with the kernel method. In 
equation~(\ref{eq:lr}), the feature vector of the $i$-th sample $\bm{x}_i$ 
can be replaced by $\bm{x}_i^\prime = (k(\bm{x}_0, \bm{x}_i), k(\bm{x}_1, \bm{x}_i), \cdots, k(\bm{x}_N, \bm{x}_i))$, where $k(\cdot)$ is the
kernel function. We use the RBF kernel: $k(\bm{x}_i, \bm{x}_j)=\exp (-\|\bm{x}_i-\bm{x}_j\|_2^2/2\sigma^2)$. The dimension of the new vector of 
features $\bm{x}^\prime$ increases from the number of features $M$ to the 
number of samples $N$. To avoid overfitting, $\ell_1$ regularization for 
$\bm{w}$ is useful. The objective function is then 
$E(\bm{w}) = \log \ell(\bm{w}) + \lambda\|\bm{w}\|_1$. The model that optimizes $E(\bm{w})$ is called the 
sparse multinomical logistic regression (SMLR).

\begin{table}
  \tbl{Confusion matrix of the LR classifier.}{%
  \begin{tabular}{cccccc}
      \hline
      \diagbox{Pred.}{Label} & N & DN & WZ & M & F\\
      \hline
      N  & 248 &  25 &  10 &  11 &  15 \\
      DN &  10 &1050 &  10 &   2 &  32 \\
      WZ &  26 & 252 & 146 &   0 &   1 \\
      M  &   5 &   2 &   0 &1026 &  10 \\
      F  &   8 & 123 &   0 &  22 & 413 \\
      \hline
      Accuracy & 0.835 & 0.723 & 0.880 & 0.967 & 0.877\\
      \hline
    \end{tabular}}\label{tab:conf_LR}
\end{table}

\begin{table}
  \tbl{Confusion matrix of the SMLR classifier.}{%
  \begin{tabular}{cccccc}
      \hline
      \diagbox{Pred.}{Label} & N & DN & WZ & M & F\\
      \hline
      N  & 241 &  20 &  11 &   9 &  12 \\
      DN &  14 &1096 &  15 &   1 &  19 \\
      WZ &  28 & 235 & 140 &   0 &   5 \\
      M  &   6 &   2 &   0 &1030 &   6 \\
      F  &   8 &  99 &   0 &  21 & 429 \\
      \hline
      Accuracy & 0.811 & 0.755 & 0.843 & 0.971 & 0.911\\
      \hline
    \end{tabular}}\label{tab:conf_SMLR}
\end{table}

We developed the LR and SMLR classifiers using samples that have more than three features. Using the same samples, GM gives the confusion matrix in 
Table~\ref{tab:conf_limit}. The hyperparameters of SMLR, $\lambda$ and 
$\sigma^2$ were determined by 10-fold cross-validation. As a result, 
the cross-validated accuracies are 0.836, 0.852, and 0.869 for the optimized 
LR, SMLR and GM classifiers, respectively. Tables~\ref{tab:conf_LR} and 
\ref{tab:conf_SMLR} are the confusion matrices for the LR and SMLR 
classifiers, respectively. The accuracy for SMLR and GM are comparable, and 
slightly higher than LR. There is no major difference in the characteristics 
of the confusion matrix for those three models: for example, all of them 
misclassify a significant part of DN as WZ. These results suggest that there 
is no significant difference in the performance of LR, SMLR, and GM. 

LR and SMLR require training data sets without missing values to classify the target. As a result, they can use limited samples in the training data. On the other hand, GM can be developed using all values in the training data, which is an advantage for GM. 

However, there are aspects of our classifier that warrant future improvement. Some of the missing values, such as distance and quiescent magnitude, are actually informative upper or lower limits. Additionally, the features $M_{\rm max}$ and $a$ should be treated as lower limits because they are derived from the magnitude at the time of discovery, when the object may still be in a rapid brightening phase. 
Implementing nonlinear classification models that can use them 
effectively will be an important direction for future research.

\subsection{Mass ejection process in novae}

\begin{figure}
 \begin{center}
  \includegraphics[width=8cm]{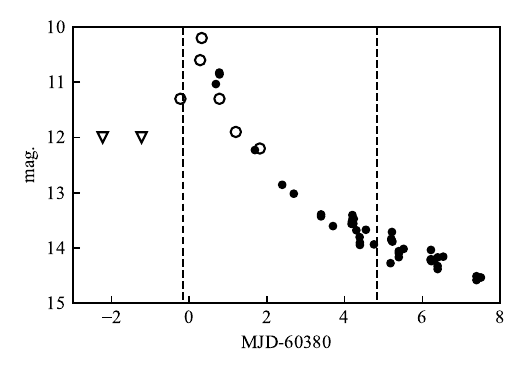} 
 \end{center}
\caption{Light curve of V4370 Oph. The data is from AAVSO. The open and 
filled circles denote visual and $V$-band magnitudes, respectively. The open 
triangles are upper-limits in the visual magnitude. The vertical dashed lines 
show the times of our spectroscopic observations.
}\label{fig:lc}
\end{figure}

V4370 Oph rapidly faded after reaching its optical maximum. 
Figure~\ref{fig:lc} illustrates the light curve of the object, sourced from 
AAVSO\footnote{https://www.aavso.org}. We removed systematic outliers of 
$V$-mag reported by specific observers. Assuming the observed maximum of 
10.4~mag at MJD$=60380.3$, we estimate the times for the object to decay by 2 
and 3 mag from maximum light to be $t_2\sim 1.6\;{\rm d}$ and $t_3\sim 3.0\;
{\rm d}$, respectively. Consequently, V4370 Oph qualifies as a very fast 
nova. The maximum magnitude rate of decline (MMRD) relationship indicates 
that the object exhibits high luminosity at maximum and a high white dwarf 
mass \citep{liv92mmrd}. Using the MMRD relation from 
\citet{del20nova}, the absolute magnitude in $V$-band, $M_V$, can be 
estimated as $M_V=-7.78-0.81\times \arctan((1.32-\log(t_2))/0.23)$, we 
obtain $M_V=-8.9$ for $t_2=1.6\;{\rm d}$. Distance to the object can be 
estimated as 9.6\;kpc using the maximum magnitude observed of 10.4 mag 
and $A_V=4.37$ \citep{sch11dust}. A quiescent counterpart with $g=21.71$ in 
the Pan-STARRS DR1 catalogue \citep{panstarrs1} suggests an absolute 
magnitude at quiescence of $M_{\rm q}=6.8$. 
This suggests a secondary star, likely a K-type main-sequence 
star or later, though this inference has significant uncertainty due to 
the large error in the distance estimation.

The times of our spectroscopic observations are indicated in 
Figure~\ref{fig:lc}: the first observation occurred about 11 hours before the 
light curve maximum, while the second was during the decline phase. The 
spectra depicted in Figures~\ref{fig:sp_all} and \ref{fig:sp_max} exhibit 
typical features of very fast novae, namely, P-Cygni profiles just before the 
maximum and broad emission lines after the maximum.

Owing to recent rapid discoveries, reports, and follow-up observations, the 
evolution of emission lines during the initial rising stage now provides new 
insights into the formation process of mass ejection in novae. 
\citet{ara15tpyx} identified highly excited lines of N\,\emissiontype{V}, 
\emissiontype{IV}, \emissiontype{III}, C\,\emissiontype{IV}, 
\emissiontype{III}, and He\,\emissiontype{II} during the rising phase of the 
recurrent nova T Pyx. They called this stage the Wolf-Rayet (WR) stage due to 
its spectral similarity. \citet{ayd23g22alz} report a comparable spectrum 
during the rising stage of the very slow nova Gaia22alz. Additionally, 
\citet{tag23v1405cas} successfully obtained spectra during the rising stage 
of V1405 Cas, revealing 
N\,\emissiontype{III} and He\,\emissiontype{II} lines, although lacking 
N\,\emissiontype{V}, \emissiontype{IV}, and C\,\emissiontype{IV} lines. These 
high-excitation lines exhibit no P Cygni profiles and have moderate 
velocities ($1000$--$2000\,{\rm km\,s^{-1}}$).

The spectra of V4370 Oph on the first day showed indications of moderately 
excited lines of Al\,\emissiontype{III}, but lacked higher excitation lines, 
suggesting a lower excitation temperature compared to the aforementioned 
samples. The lack of highly excited lines and presence of 
Al\,\emissiontype{III} lines resemble the spectrum of V1405 Cas taken after 
the initial rapid rise and in the early stage of the pre-maximum halt (at 
53.53 hours after its discovery) \citep{tag23v1405cas}.

According to the standard model of nova eruptions (e.g., \cite{bod08nova}), 
the envelope surrounding a white dwarf begins to expand when the photospheric 
luminosity reaches the Eddington limit. This is the so-called fireball stage, 
during which a soft X-ray flash was recently detected in YZ~Ret 
\citep{kon22yzret, kat22yzret}. Initially, the photospheric temperature is 
expected to be $\lesssim 10^6,{\rm K}$, decreasing gradually while 
maintaining a constant luminosity. The physical conditions likely resemble 
those of WR stars, with a photospheric temperature in the range of a few 
$10^5\,{\rm K}$ shortly after the onset of expansion. The high-excitation 
lines observed in T Pyx and Gaia22alz would likely originate under such 
conditions. In this scenario, our observation of V4370 Oph is definitely 
capturing a subsequent stage when the temperature decreases moderately, but 
still before reaching the minimum.

The decrease in photospheric temperature can also account for the rapid 
deepening of the absorption component of H$\alpha$, as shown in 
Figure~\ref{fig:Halpha_var}. The variation in line profile was predominantly 
observed in the absorption component rather than the entire profile of the 
emission line. Therefore, it is likely attributed to changes not in the 
region where emission line photons are generated, but rather in the 
photosphere. Balmer absorption lines are expected to be weak in the condition 
that high-excitation lines are formed and hydrogen gas is fully ionized 
during the early WR stage. We propose that the reduction in photospheric 
temperature toward $\sim 10^4\,{\rm K}$ would consequently lead to the 
strengthening of the absorption component.

\citet{yam10usco} also successfully obtained the time-series spectra of U Sco 
during its initial rising stage. Although the spectra exhibited no high-
excited lines, rapid evolution of the high-velocity component was observed. 
The Balmer and He\,\emissiontype{I} emission lines showed a triple-peak 
profile, with the blue and red peaks at $\sim 3000\,{\rm km\,s^{-1}}$ 
developing more rapidly than the central peak within 1.7~hours. These high-
velocity components became dominant, and the emission lines exhibited a box-
like profile one day after the initial observation. Furthermore, 
\citet{ayd20novae} reported early spectra of 12 novae and noted that the 
delayed growth of the high-velocity component is a common feature among novae.

The delayed appearance of the high-velocity component in emission lines 
implies a postponed launch of the optically thick wind \citep{kat94wind}. 
This wind, expected to be formed due to acceleration by the iron opacity bump 
at $\sim 1.5\times 10^5\,{\rm K}$ in the optically thick envelope, is 
proposed to experience a delay in its evolution due to binary interaction 
\citep{she22novabin}. According to \citet{she22novabin}, a slow equatorial 
outflow is first driven by the binary interaction, and then the fast wind is 
launched after the acceleration region recedes within the WD Roche radius.

Our observations revealed no indication of a high-velocity component in the 
spectra of V4370 Oph on the first day, whereas a box-like profile was 
observed on the fifth day. Combined with the short $t_2$ and $t_3$ 
values, V4370 Oph potentially resembles U Sco, which also exhibited a
box-like profile on the second day. The stage at which we observed V4370 Oph 
on the first day may correspond to just before the stage observed in U Sco. 
In summary, we propose that the spectra of V4370 Oph taken on the first day 
correspond to the termination of the WR stage, during which the 
envelope's photosphere expands and the optically thick wind has yet to be 
launched.

However, the FWHM of the emission lines was already moderately high, exceeding
$1000\,{\rm km\,s^{-1}}$, on the first day. Furthermore, T Pyx and V1405 Cas
also exhibited high velocities that exceeded $1000\,{\rm km\,s^{-1}}$ during 
the initial rising stage (\cite{ara15tpyx, tag23v1405cas}). If the optically 
thick wind had not yet formed, that is, the acceleration at the iron opacity 
bump had not been effective, another acceleration mechanism may be necessary 
to explain the initial outflow. The decay energy of $\beta$-unstable 
nuclei (mainly $^{13}$N and $^{15}$O) can directly accelerate the envelope, 
although this would likely occur only in the very early phase when the 
polluted area is close to the envelope surface (\cite{sta16nova, 
she22novabin}).

The scenario described above requires validation through further data, as 
previous samples display significant diversity in their light curves. Both 
V4370 Oph and U Sco are classified as very fast novae undergoing rapid rises. 
V1405 Cas also experienced a rapid rise; however, the object displayed a 
prolonged plateau with short-lived flares over 200~d \citep{tag23v1405cas}. 
On the contrary, T Pyx first exhibited a rapid rise, followed by a gradual 
rise, and finally a slow decline \citep{cho14tpyx}. Gaia22alz is 
characterized as a very slow nova, with high excitation lines observed 49 
days after the eruption onset, still during the initial rising stage 
\citep{ayd23g22alz}. It is not evident that the evolution of the emission 
lines is identical, as only the rising and decline time-scales differ. 
Additionally, geometric effects may play an important role in determining the 
velocity and profile variations, as novae have shown signs of aspherical 
ejecta (e.g., \cite{cho14v959mon}). Therefore, more data covering a wide 
range of binary parameters are necessary to establish the observational 
features of the initial rising stage. We have demonstrated that Smart Kanata 
effectively contributes to this study through a case study of V4370 Oph.

\section{Summary}

We have developed Smart Kanata, an autonomous decision-making system for 
prompt follow-up observations of CVs using the 1.5-m Kanata 
telescope. The system selects the optimal observation mode with 
the highest MI among three possible options. This MI is calculated based on 
the class probabilities estimated from the GM classifier of variable star 
types and the likelihood functions constructed from known samples. During its 
$\sim 300$ days of operation, our system demonstrated its capability to make 
appropriate decisions for follow-up observations across 21 samples. It 
successfully conducted automated spectroscopic observations for the nova 
V4370 Oph and the microlensing event TCP J17440698$-$2125195. Notably, in the 
time-series spectra of V4370 Oph, we observed a rapid deepening of the 
absorption component of the H$\alpha$ line. These findings highlight the 
potential of Smart Kanata as a valuable tool for investigating the early 
phases of transient events in CVs.

Future developments for Smart Kanata include expanding the number of online 
platforms it monitors. These may include existing platforms such as Gaia 
Alerts \citep{hod21gaia_alert} and, in the future, transient events 
discovered by LSST \citep{ive19lsst}. Additionally, updating the 
classification model to effectively handle missing values in training data, 
as well as upper and lower limits, will be a critical improvement.

\begin{ack}
We acknowledge with thanks the variable star observations from the AAVSO 
International Database contributed by observers worldwide and used in this 
research. We would like to thank the transient survey projects ATLAS, Gaia 
Alert, XOSS, MASTER, ASAS-SN, BraTS, as well as the databases TNS and CBAT,  
for providing prompt information about transients, which was essential in 
conducting this study.
\end{ack}

\section*{Funding}
This work was supported by a Kakenhi Grant-in-Aid (No. 21K03616) from  
Japan Society for the Promotion of Science (JSPS). 

\appendix 
\section{Multivariate normal distributions of training samples for the GM classifier}

As discussed in Section 2.3, we calculated the means and covariances of the 
training samples to construct a GM classifier. Table~\ref{tab:gm_mean} lists 
the means, while Tables~\ref{tab:gm_cov_N}, \ref{tab:gm_cov_DN}, 
\ref{tab:gm_cov_WZ}, \ref{tab:gm_cov_M}, and \ref{tab:gm_cov_F} present the 
variance-covariance matrices for classes N, DN, WZ, M, and F, respectively. 
We note that class M has a negative mean for $a$, even though it is expected 
to be positive. This discrepancy arises because, as mentioned in Section 2.3, 
$M_{\rm max}$ is derived from relatively blue bands ($V$, $CV$, $g$, and 
$pg$) in the VSX catalog, while $M_{\rm qui}$ is derived from the broad Gaia 
$G$-band. As a result, objects tend to be brighter in $M_{\rm qui}$ than in 
$M_{\rm max}$.

\begin{table*}
  \tbl{The means of the features of the training samples used for classification.}{%
  \begin{tabular}{crrrrrrrr}
      \hline
      Class & $\ell$ & $b$ & $M_{\rm max}$ & $M_{\rm qui}$ & $a$ & $g-r$  & $r-i$ & $i-z$\\
      \hline
      N & $4.395$ & $-1.452$ & $-4.201$ & $5.322$ & $9.716$ & $0.679$ & $0.349$ & $0.281$ \\
      DN & $20.562$ & $0.734$ & $5.171$ & $8.668$ & $3.730$ & $0.239$ & $0.175$ & $0.232$ \\
      WZ & $12.800$ & $0.683$ & $4.260$ & $11.600$ & $7.300$ & $0.060$ & $-0.033$ & $0.230$ \\
      M & $7.728$ & $-1.477$ & $1.117$ & $-0.394$ & $-1.536$ & $3.326$ & $2.277$ & $1.471$ \\
      F & $27.837$ & $-3.261$ & $7.822$ & $9.377$ & $1.815$ & $1.165$ & $1.245$ & $0.646$ \\
      \hline
    \end{tabular}}\label{tab:gm_mean}
\end{table*}

\begin{table*}
  \tbl{The variance-covariance matrix of the features of the training samples of class N used for classification.}{%
  \begin{tabular}{crrrrrrrr}
      \hline
       & $\ell$ & $b$ & $M_{\rm max}$ & $M_{\rm qui}$ & $a$ & $g-r$  & $r-i$ & $i-z$\\
      \hline
$\ell$       & $3170.073$ & $34.244$  & $-38.674$ & $-4.629$ & $17.302$ & $0.629$  & $4.621$  & $2.413$  \\
$b$       & $34.244$  & $110.677$ & $4.992$   & $1.184$  & $-2.185$ & $-0.109$ & $-0.044$ & $-0.036$ \\
$M_{\rm max}$ & $-38.674$ & $4.992$   & $8.928$   & $5.405$  & $-3.523$ & $0.600$  & $-0.185$ & $0.082$  \\
$M_{\rm qui}$ & $-4.629$  & $1.184$   & $5.405$   & $11.549$ & $6.144$  & $-0.315$ & $-0.409$ & $-0.317$ \\
$a$      & $17.302$  & $-2.185$  & $-3.523$  & $6.144$  & $8.134$  & $-0.847$ & $-0.246$ & $-0.348$ \\
$g-r$     & $0.629$   & $-0.109$  & $0.600$   & $-0.315$ & $-0.847$ & $0.582$  & $0.128$  & $0.123$  \\
$r-i$     & $4.621$   & $-0.044$  & $-0.185$  & $-0.409$ & $-0.246$ & $0.128$  & $0.346$  & $0.084$  \\
$i-z$     & $2.413$   & $-0.036$  & $0.082$   & $-0.317$ & $-0.348$ & $0.123$  & $0.084$  & $0.204$  \\
      \hline
    \end{tabular}}\label{tab:gm_cov_N}
\end{table*}

\begin{table*}
  \tbl{The variance-covariance matrix of the features of the training samples of class DN used for classification.}{%
  \begin{tabular}{crrrrrrrr}
      \hline
       & $\ell$ & $b$ & $M_{\rm max}$ & $M_{\rm qui}$ & $a$ & $g-r$  & $r-i$ & $i-z$\\
      \hline
$\ell$       & $9558.629$ & $-68.459$ & $9.949$   & $-9.002$ & $-20.457$ & $3.363$  & $0.883$  & $2.055$  \\
$b$       & $-68.459$  & $737.777$ & $0.559$   & $-1.552$ & $-0.599$  & $-0.292$ & $-0.564$ & $-0.696$ \\
$M_{\rm max}$ & $9.949$   & $0.559$   & $1.664$   & $1.425$  & $-0.238$  & $-0.000$ & $0.023$  & $0.012$  \\
$M_{\rm qui}$ & $-9.002$  & $-1.552$  & $1.425$   & $4.806$  & $3.380$   & $-0.056$ & $-0.004$ & $0.118$  \\
$a$      & $-20.457$ & $-0.599$  & $-0.238$  & $3.380$  & $4.344$   & $-0.090$ & $-0.021$ & $0.106$  \\
$g-r$     & $3.363$  & $-0.292$  & $-0.000$  & $-0.056$ & $-0.090$  & $0.329$  & $-0.035$ & $-0.013$ \\
$r-i$     & $0.883$  & $-0.564$  & $0.023$   & $-0.004$ & $-0.021$  & $-0.035$ & $0.219$  & $-0.048$ \\
$i-z$     & $2.055$  & $-0.696$  & $0.012$   & $0.118$  & $0.106$   & $-0.013$ & $-0.048$ & $0.253$  \\
      \hline
    \end{tabular}}\label{tab:gm_cov_DN}
\end{table*}

\begin{table*}
  \tbl{The variance-covariance matrix of the features of the training samples of class WZ used for classification.}{%
  \begin{tabular}{crrrrrrrr}
      \hline
       & $\ell$ & $b$ & $M_{\rm max}$ & $M_{\rm qui}$ & $a$ & $g-r$  & $r-i$ & $i-z$\\
      \hline
$\ell$       & $10300.169$ & $-159.818$ & $-2.867$   & $3.619$  & $-2.652$ & $0.161$  & $-1.338$ & $0.682$  \\
$b$       & $-159.818$  & $1175.438$ & $0.665$   & $-0.978$ & $-0.009$ & $-0.443$ & $-1.393$ & $-3.670$ \\
$M_{\rm max}$ & $-2.867$   & $0.665$   & $0.876$   & $0.241$  & $-0.641$ & $0.001$  & $-0.010$ & $0.042$  \\
$M_{\rm qui}$ & $3.619$   & $-0.978$  & $0.241$   & $0.453$  & $0.212$  & $-0.018$ & $-0.011$ & $-0.039$ \\
$a$      & $-2.652$   & $-0.009$  & $-0.641$  & $0.212$  & $0.894$  & $-0.002$ & $-0.013$ & $-0.039$ \\
$g-r$     & $0.161$    & $-0.443$  & $0.001$   & $-0.018$ & $-0.002$ & $0.044$  & $0.011$  & $0.028$  \\
$r-i$     & $-1.338$   & $-1.393$  & $-0.010$  & $-0.011$ & $-0.013$ & $0.011$  & $0.088$  & $0.018$  \\
$i-z$     & $0.682$    & $-3.670$  & $0.042$   & $-0.039$ & $-0.039$ & $0.028$  & $0.018$  & $0.270$  \\
      \hline
    \end{tabular}}\label{tab:gm_cov_WZ}
\end{table*}

\begin{table*}
  \tbl{The variance-covariance matrix of the features of the training samples of class M used for classification.}{%
  \begin{tabular}{crrrrrrrr}
      \hline
       & $\ell$ & $b$ & $M_{\rm max}$ & $M_{\rm qui}$ & $a$ & $g-r$  & $r-i$ & $i-z$\\
      \hline
$\ell$       & $7404.059$ & $112.985$  & $59.471$  & $13.047$ & $-45.318$ & $10.062$ & $5.469$  & $2.198$  \\
$b$       & $112.985$  & $192.225$  & $2.933$   & $-0.729$ & $-3.972$  & $0.174$  & $0.010$  & $0.528$  \\
$M_{\rm max}$ & $59.471$   & $2.933$   & $7.396$   & $2.052$  & $-5.344$ & $0.775$  & $0.569$  & $0.275$  \\
$M_{\rm qui}$ & $13.047$  & $-0.729$  & $2.052$   & $1.989$  & $-0.063$ & $0.184$  & $0.332$  & $0.118$  \\
$a$      & $-45.318$  & $-3.972$  & $-5.344$  & $-0.063$ & $5.279$  & $-0.557$ & $-0.264$ & $-0.140$ \\
$g-r$     & $10.062$   & $0.174$   & $0.775$   & $0.184$  & $-0.557$ & $2.506$  & $-0.556$ & $-0.030$ \\
$r-i$     & $5.469$    & $0.010$   & $0.569$   & $0.332$  & $-0.264$ & $-0.556$ & $1.303$  & $0.041$  \\
$i-z$     & $2.198$    & $0.528$   & $0.275$   & $0.118$  & $-0.140$ & $-0.030$ & $0.041$  & $0.231$  \\
      \hline
    \end{tabular}}\label{tab:gm_cov_M}
\end{table*}

\begin{table*}
  \tbl{The variance-covariance matrix of the features of the training samples of class F used for classification.}{%
  \begin{tabular}{crrrrrrrr}
      \hline
       & $\ell$ & $b$ & $M_{\rm max}$ & $M_{\rm qui}$ & $a$ & $g-r$  & $r-i$ & $i-z$\\
      \hline
$\ell$       & $14952.701$ & $-821.461$ & $-24.930$ & $-66.631$ & $-42.921$ & $-4.859$ & $-15.800$ & $-8.642$ \\
$b$       & $-821.461$  & $1216.280$ & $5.908$  & $9.737$  & $9.813$  & $1.150$  & $3.017$  & $1.538$  \\
$M_{\rm max}$ & $-24.930$  & $5.908$   & $5.896$  & $6.024$  & $0.128$  & $0.443$  & $1.080$  & $0.511$  \\
$M_{\rm qui}$ & $-66.631$  & $9.737$   & $6.024$  & $9.210$  & $3.186$  & $0.626$  & $1.723$  & $0.850$  \\
$a$      & $-42.921$  & $9.813$   & $0.128$  & $3.186$  & $4.288$  & $0.174$  & $0.611$  & $0.397$  \\
$g-r$     & $-4.859$   & $1.150$   & $0.443$  & $0.626$  & $0.174$  & $0.130$  & $0.093$  & $0.061$  \\
$r-i$     & $-15.800$  & $3.017$   & $1.080$  & $1.723$  & $0.611$  & $0.093$  & $0.421$  & $0.142$  \\
$i-z$     & $-8.642$   & $1.538$   & $0.511$  & $0.850$  & $0.397$  & $0.061$  & $0.142$  & $0.138$  \\
      \hline
    \end{tabular}}\label{tab:gm_cov_F}
\end{table*}

\section{Classifier for the weather condition}

As discussed in Section 2.4, we automatically evaluate weather conditions 
using images captured by the all-sky monitor camera at our observatory. In 
this section, we provide details on the CNN classifier developed for this 
purpose.

\begin{figure}
 \begin{center}
  \includegraphics[width=8cm]{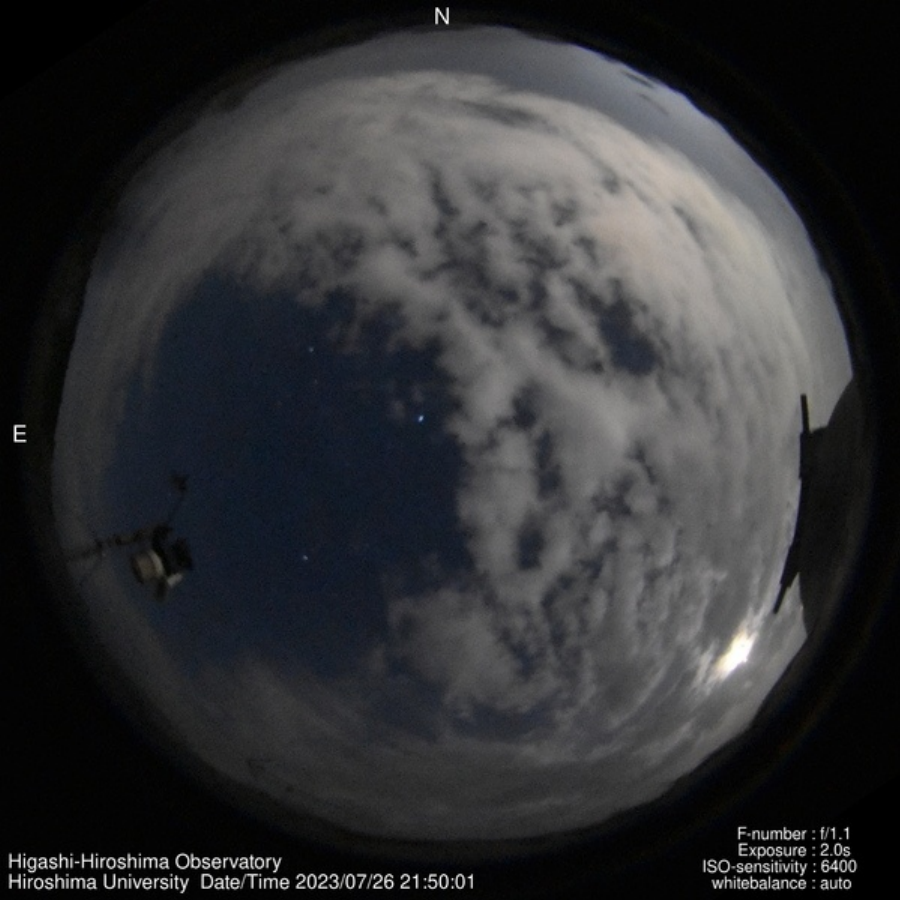} 
 \end{center}
\caption{An example of the images taken using the sky-monitor camera at the Higashi-Hiroshima Observatory.
}\label{fig:skymon}
\end{figure}

The training data set comprises 70659 nighttime sky images captured between 
15 November 2021 and 16 August 2023. These original RGB color images, each 
with dimensions of $600\times 600$ pixels, were taken using an optical 
camera. An example of these images is presented in Figure~\ref{fig:skymon}. 
To preprocess the images, we initially trim the outer sections, creating 
$400\times 400$ pixel images that only depict the sky. Subsequently, these 
images were resized to $32\times 32$ pixels. Consequently, the CNN input 
consists of three-color images with dimensions of $32\times32\times3$. 
Each image is labeled either "good" or "bad." The "good" label indicates
minimal cloud coverage suitable for observation, while the "bad" label 
denotes significant cloud presence that prevents observation. The labeling 
process was performed manually, resulting in a dataset of 41419 "good" and 
29240 "bad" images.

\begin{table*}
  \tbl{CNN architecture.}{%
  \begin{tabular}{ccccc}
      \hline
      Layer type & Filter/Pooling size & No. of units & Output size & No. of parameters\\
      \hline
      Input & --- & --- & $32\times 32\times 3$ & --- \\
      Convolution \#1 & $3\times 3\times 3$ & 32 & $30\times 30\times 32$ & 896 \\
      Max pooling \#1 & $2\times 2$ & --- & $15\times 15\times 32$ & 0 \\
      Convolution \#2 & $2\times 2\times 32$ & 64 & $14\times 14\times 64$ & 8256 \\
      Max pooling \#2 & $2\times 2$ & --- & $7\times 7\times 64$ & 0 \\
      Fully connected \#1 & --- & 128 & 128 & 401536\\
      Fully connected \#2 & --- & 1 & 1 & 129\\
      \hline
    \end{tabular}}\label{tab:cnn}
\end{table*}

Table~\ref{tab:cnn} provides an overview of the CNN model used for this
classification task. The model consists of two convolution layers followed by 
two fully connected layers. ReLU activation functions are applied to all
layers, except the second fully connected layer, which plays the role of the 
output layer and employs the sigmoid function. To enhance the model's 
robustness, dropout layers were incorporated after the second convolution 
layer and the first fully connected layer, both with a dropout rate of 0.5.

The model was optimized using the training data. The cost function used was 
cross-entropy and the optimization algorithm used was ADAM \citep{adam}. A 
batch size of 256 was applied and 20\% of the training data was reserved for 
validation purposes. The optimization process completed when there was no 
improvement in the accuracy of the validation data for 10 consecutive steps.

\begin{table}
  \tbl{Confusion matrix of the weather classifier.}{%
  \begin{tabular}{ccc}
      \hline
      \diagbox{Pred.}{Label} & good & bad \\
      \hline
      Good  & 13664 &  277 \\
            &(40803) &(288) \\
      Bad   & 321 & 10009 \\
            &(616) & (28952) \\
      \hline
      Accuracy & 0.977 & 0.973 \\
               & (0.985) & (0.990)\\
      \hline
  \end{tabular}}\label{tab:cnn_mat}
\end{table}

We acquired additional 24271 sky images from 12 October 2023 to 12 September 
2024, to evaluate the generalized performance of the optimized model. 
Table~\ref{tab:cnn_mat} presents the confusion matrix for this test data 
sets, with values in parentheses corresponding to those for the training 
data. The accuracy for the test data is 0.975, comparable to the training 
data accuracy of 0.987. The confusion matrix for the test data exhibits 
similar features to that of the training data, suggesting no significant 
overfitting to the training dataset. The classifier accuracy is high enough 
for the automated observation of Smart Kanata.


\end{document}